%% file: main.tex
\documentclass[11pt]{article}

\usepackage[utf8]{inputenc}
\usepackage[T1]{fontenc}
\usepackage{lmodern}
\usepackage{geometry}
\usepackage{graphicx}
\usepackage{booktabs}
\usepackage{array}
\usepackage{tabularx}
\usepackage{xcolor}
\usepackage{listings}
\usepackage{authblk}
\usepackage[numbers,sort&compress]{natbib}
\usepackage{url}
\usepackage{xurl}
\usepackage{amsmath}
\usepackage[section]{placeins}
\usepackage{float}
\usepackage{hyperref}

\definecolor{darkblue}{RGB}{0,0,120}
\hypersetup{colorlinks=true, citecolor=darkblue, linkcolor=darkblue, urlcolor=darkblue,
  pdftitle={How Deterministic SAST and AI Security Agents Divide the Work: Coverage, Harness Policy and Cost in Snyk VulnBench 2.0},
  pdfauthor={Liran Tal, Chris Liu, Charlotte Fereday, Sebastian Roth, Brian Clark},
  pdfsubject={Benchmarking deterministic SAST and AI security agents},
  pdfkeywords={SAST, AI security agents, vulnerability detection, benchmark, Snyk Code, Claude Code, Codex Security}}

\newcommand{\secrev}{\texttt{/security-review}}

\title{How Deterministic SAST and AI Security Agents Divide the Work:\\Coverage, Harness Policy and Cost in Snyk VulnBench 2.0}
\author[1]{Liran Tal}
\author[1]{Chris Liu}
\author[1]{Charlotte Fereday}
\author[1]{Sebastian Roth}
\author[1]{Brian Clark}
\affil[1]{Snyk\\\texttt{\{liran,xian.liu,charlotte.fereday,sebastian.roth,brian.clark\}@snyk.io}}
\date{October 2026}

\begin{document}

\maketitle

\input{sections/00-abstract}

\input{sections/01-introduction}

\input{sections/02-method}

\input{sections/03-complementarity}

\input{sections/04-harness-policy}

\input{sections/05-efficiency}

\input{sections/06-robustness}

\input{sections/07-threats}

\input{sections/08-related-work}

\input{sections/09-future-work}

\input{sections/10-conclusion}

\input{sections/11-availability}

\bibliographystyle{plainnat}
\bibliography{references}

\appendix
\input{sections/A-corpus}

\input{sections/B-configurations}

\input{sections/C-supplementary}

\end{document}

%% file: sections/00-abstract.tex
\begin{abstract}
Security teams now point AI coding agents and AI security harnesses at their applications. We ask what these engines find that deterministic static application security testing (SAST) does not, and what they miss that SAST catches. VulnBench~2.0 is a benchmark of 108 hand-curated, attacker-reachable vulnerabilities in 20 open-source web applications written in Go, JavaScript/TypeScript, Python and Java; 62 of the vulnerabilities flow across more than one file. A finding is credited only if it locates the vulnerability's attacker-controlled source and its dangerous sink, each within two lines. We evaluate nine productized configurations under the same blind protocol: the Snyk Code static analyzer, Claude Code's \secrev{} with three model and effort settings (one with the Snyk MCP server), Codex Security with GPT-5.6 Luna, GPT-5.6 Terra and GPT-5.6 Sol, and the DeepSec scanner with two model backends. Snyk Code reached 82.4\% macro recall (89 of 108) and Claude Code with Opus~5 at xhigh effort 79.3\% (85 of 108); they disagree on only 28 vulnerabilities, split 16 to 12, a gap this corpus is too small to tell apart from chance. In this study neither engine was sufficient alone. Snyk Code contributed coverage no AI configuration reproduced, the AI configurations contributed cross-file and logic findings Snyk Code missed, and together the two reached 101 of 108, the ceiling of all seven comparable configurations combined. The AI configurations also reported at least seven real vulnerabilities in scored classes that the reference set did not contain and Snyk Code did not report. What an AI engine misses depends on its harness as much as on its model: Claude Code's built-in review policy excludes or restricts three vulnerability classes, and in a supplementary run on 19 of the 20 applications the same model without that policy matched 16 of 20 vulnerabilities in those classes instead of 9; Codex Security with GPT-5.6 Sol located 80 vulnerabilities at file level but 34 at the line level; DeepSec's model-free file selection kept a file from only 56 of the 108 vulnerable flows. Coverage and cost were only loosely related. The harness, fixtures, reference sets and all run data are public.
\end{abstract}

%% file: sections/01-introduction.tex
\section{Introduction}
\label{sec:intro}

AI coding agents now review code before people do, and several vendors ship agentic harnesses built specifically for security review. Security teams therefore face a practical choice that earlier benchmarks answer only in part: when an AI engine is pointed at a real application, what does it find that a deterministic static analyzer does not, and what does it miss that the analyzer catches? We lead with recall, because the question practitioners ask of a scanner is whether it finds the vulnerabilities that are actually there; precision and cost are reported beside it.

Head-to-head comparisons of SAST tools and language models already exist. They report that language models detect more vulnerabilities at the cost of more false positives, and that combining the two families helps~\citep{zhou2024sastllm,gnieciak2025llmsast}. Most of them call language models directly, rather than scoring the packaged security tools that teams actually run, such as an agent's built-in review command or a vendor's scanning harness; credit a finding for naming the right vulnerability type or file, and rely on ground truth that is either derived from a tool or limited to one language. Our own predecessor, VulnBench JS~1.0~\citep{tal2026vulnbench1}, measured how repeatably Claude Code found the vulnerabilities in ten small JavaScript applications. It used Snyk Code's findings as the reference set and credited a match by vulnerability type alone, and it named the next steps: application-scale fixtures, an independent ground truth, and stricter scoring.

VulnBench~2.0 takes those steps. It changes four things.
\begin{itemize}
  \item \textbf{Application-scale fixtures.} Twenty open-source web applications in four language families, with 108 vulnerabilities, 62 of which flow across more than one file.
  \item \textbf{Curated, attacker-reachable ground truth.} The reference set is not derived from any tool's findings: every vulnerability was curated as reachable by an outside attacker, with an annotated source-to-sink flow and an exploit proof of concept verified by the authors. A vulnerability class an engine does not report is therefore a real miss, not noise reduction.
  \item \textbf{Endpoint-aware scoring.} A finding counts only if it locates the attacker-controlled source and the dangerous sink of the curated flow, each within two lines, with one finding credited to at most one vulnerability.
  \item \textbf{Productized harnesses.} We evaluate the configurations practitioners run: the Snyk Code analyzer, Claude Code's built-in \secrev{} command, OpenAI's Codex Security, and the DeepSec scanner. We attribute every result to the complete harness and model configuration, never to a model alone.
\end{itemize}

Our answer is that, in this study, neither kind of engine was sufficient alone, and that the two divide the work. Deterministic SAST contributed coverage that no AI configuration reproduced; the AI configurations contributed cross-file and logic findings that SAST missed; together, Snyk Code and one AI review reached the ceiling of what all comparable configurations found. What an AI engine misses depends heavily on its harness, not only on its model. Four findings support this.

\paragraph{Complementarity (Section~\ref{sec:complementarity}).} Snyk Code matched 82.4\% of the reference set (89 of 108) and Claude Code \secrev{} with Opus~5 at xhigh effort 79.3\% (85 of 108). They disagree on only 28 vulnerabilities, split 16 to 12, and a corpus of this size cannot tell a gap that small from chance. Their union matches 101 of 108, the same as all seven comparable configurations combined. They contribute different vulnerabilities: Snyk Code adds 16 to the agent, mostly within single files and in classes such as open redirect, while the agent adds 12, seven of them cross-file, including the authorization classes. Snyk Code is the only one of the seven endpoint-aware configurations with unique coverage. Outside the reference set, the AI configurations reported at least seven real vulnerabilities in scored classes that the reference set did not contain and Snyk Code did not report.

\paragraph{Harness policy (Section~\ref{sec:harness}).} Claude Code's \secrev{} excludes regular-expression denial of service and reports prototype pollution and open redirect only at extremely high confidence. On the 20 reference vulnerabilities in those classes, Snyk Code matched 18 and Opus~5 xhigh 9; on the 81 others Opus~5 xhigh matched 71 and Snyk Code 67. In a supplementary run on 19 of the 20 applications, the same model at the same effort, given the benchmark's default audit prompt in place of \secrev{}, matched 16 of the 20. Codex Security with GPT-5.6 Sol located 80 vulnerabilities at file level but 34 under the two-line rule, typically pointing above the dangerous call. DeepSec's model-free scan stage queued a file from only 56 of the 108 flows before any model ran. The model still matters inside a harness: under the same harness, effort and prompt, Opus~5 matched 14 vulnerabilities that Sonnet~5 missed, against 2 the other way.

\paragraph{Efficiency (Section~\ref{sec:efficiency}).} Snyk Code scanned all 20 applications in 8 minutes; the comparable AI configurations took between 2.4 and 8.9 hours. A more expensive model did not buy more coverage: within Codex Security, GPT-5.6 Sol cost 35.0 times as much as GPT-5.6 Luna and matched 11 fewer vulnerabilities.

\paragraph{Robustness (Section~\ref{sec:robustness}).} Snyk Code and Opus~5 xhigh remain the top two under all 20 alternative scoring policies we tested.

All authors are employees of Snyk, which develops Snyk Code; Section~\ref{sec:threats-gt} describes how the corpus and reference set were built and the biases that follow.

\paragraph{Contributions.}
\begin{enumerate}
  \item A public benchmark of 20 applications with curated, attacker-reachable source-to-sink ground truth, each vulnerability backed by an author-verified exploit proof of concept (447 authored flow locations), together with a deterministic scorer and the complete run archive.
  \item To our knowledge, the first measurement of SAST-versus-agent overlap on endpoint-matched ground truth, including a split by whether the flow crosses a file boundary.
  \item To our knowledge, the first benchmark evaluation on curated, endpoint-annotated ground truth of Claude Code \secrev{}, Codex Security and DeepSec as security scanners, including the offset between a harness's reported line and the curated sink, and a recall ceiling set by a harness before its model runs. Concurrent work evaluates Codex Security on real vulnerability-introducing code changes~\citep{yang2026secprbench}.
  \item A sensitivity analysis of how line tolerance and matching policy change localization scores, and a prompt-swap control, with outcome thresholds fixed before its per-class analysis, that tests the harness-policy reading.
\end{enumerate}

Section~\ref{sec:method} describes the benchmark and method, Sections~\ref{sec:complementarity}--\ref{sec:robustness} report the findings, and Section~\ref{sec:threats} discusses threats to validity. Related work, future work and data availability follow.

%% file: sections/02-method.tex
\section{Benchmark and method}
\label{sec:method}

\subsection{Corpus}
\label{sec:corpus}

The corpus consists of 20 open-source web applications and services. Sixteen are third-party projects whose upstream repositories are recorded in the fixture metadata; four were written by Snyk employees for earlier work unrelated to this benchmark and later reused here: two Java applications, coffeeshop and vinyl-marketplace, and two intentionally vulnerable demonstration applications by the first author, keystonebank and sassyreg. It contains 108 reference vulnerabilities, between 2 and 10 per application (median 5). For 62 of them the curated flow from source to sink spans more than one file; the other 46 stay within a single file. The applications cover four language families (Table~\ref{tab:corpus}) and the vulnerabilities fall into 14 classes, led by cross-site scripting (19), path traversal (16), command injection (13) and SQL injection (13). Each vulnerability is annotated with its full code flow: across the corpus there are 447 authored flow locations, 131 marked as sources, 118 as sinks and 198 as unlabeled intermediate steps, about 4.1 per vulnerability. Appendix~\ref{app:corpus} gives the per-application details, the class distribution and the flow lengths.

\begin{table}[H]
\centering
\caption{Corpus summary by language family. An application with several languages is counted under its first-listed language; JavaScript and TypeScript are combined.}
\label{tab:corpus}
\begin{tabular}{lrrrr}
\toprule
Language family & Applications & Vulnerabilities & Cross-file & Same-file \\
\midrule
Go & 6 & 32 & 25 & 7 \\
JavaScript / TypeScript & 7 & 30 & 10 & 20 \\
Python & 5 & 34 & 21 & 13 \\
Java & 2 & 12 & 6 & 6 \\
\midrule
Total & 20 & 108 & 62 & 46 \\
\bottomrule
\end{tabular}
\end{table}

\subsection{Reference set}
\label{sec:reference}

The reference vulnerabilities are a mix of two origins: vulnerabilities already present in intentionally vulnerable demonstration applications, and vulnerabilities the authors seeded with an AI coding assistant (Cursor), using OpenAI models, including GPT-5.6 Luna, GPT-5.6 Terra and GPT-5.6 Sol, and Anthropic models. The origin of each vulnerability and which model seeded which application were not recorded. Every vulnerability was then curated by the authors with an annotated source-to-sink flow and an exploit proof of concept verified by the authors. The reference set is not derived from any tool's findings; Section~\ref{sec:threats-gt} discusses how Snyk Code and the seeding models were involved and what that could bias. Every entry is \emph{attacker-reachable}, meaning that input an outside attacker controls (a request parameter or body, uploaded data, stored user content, or a response from an attacker-influenced service) reaches a dangerous operation without being neutralized. For each vulnerability the reference set records its class and any conservative alternative class names, the source and sink locations (several alternative anchors are allowed for either role), the intermediate locations, and whether the flow crosses a file boundary. The reference set used for scoring is the frozen snapshot stored in the run manifest; it is identical to the public answer files on every field the scorer reads for 19 of the 20 applications; the iguana answer file now carries the correction described in Section~\ref{sec:threats-other}.

Because the evaluated configurations are blind to the reference set, a vulnerability class that an engine does not report is a real miss rather than noise reduction. Conversely, a reported finding that matches no reference vulnerability is \emph{outside the reference set}; it is not necessarily wrong, and we examine a subset of such findings in Section~\ref{sec:outside}. One known defect in the reference set is disclosed in Section~\ref{sec:threats}.

\subsection{Configurations}
\label{sec:configs}

We evaluate nine configurations (Table~\ref{tab:configs}). Each is a complete harness and model combination, and we attribute every result to that combination.

\begin{itemize}
  \item \textbf{Snyk Code} runs as the deterministic SAST baseline through its command-line interface, \texttt{snyk code test}, with no prompt.
  \item \textbf{Claude Code \secrev{}} is Claude Code's built-in security-review command, driven through the Claude Agent SDK. Three configurations vary the model and reasoning effort: Opus~5 at xhigh effort, Sonnet~5 at xhigh effort, and Opus~5 at medium effort with the Snyk MCP server attached. In that last configuration the agent is required to call \texttt{snyk\_code\_scan} exactly once and, by construction, receives Snyk Code's findings for the application (the MCP responses themselves were not recorded); it does not receive coverage information about what Snyk Code analyzed.
  \item \textbf{Codex Security} is OpenAI's security scanner, run in its standard, report-only, full-repository mode with GPT-5.6 Luna, GPT-5.6 Terra or GPT-5.6 Sol at xhigh effort.
  \item \textbf{DeepSec} is an agentic security scanner that runs a model-free scan stage followed by model-driven processing and export. It runs with an Opus~5 or a GPT-5.6 Sol backend, 150 turns per batch, batches of five files, one batch at a time.
\end{itemize}

Claude Code configurations receive the benchmark's system prompt, with the user turn replaced by \secrev{}; Codex Security and DeepSec run on their own built-in scan instructions, and the benchmark prompt is not passed to them. Appendix~\ref{app:configs} reproduces the benchmark's prompts and records every version and how it is known. We paraphrase, rather than reproduce, the vendors' built-in prompts. Figures abbreviate configuration names, for example ``Opus 5 med + MCP'' for Claude Code \secrev{} with Opus~5 at medium effort and the Snyk MCP server, and ``CxSec'' for Codex Security.

\begin{table}[H]
\centering
\small
\caption{The nine configurations. Versions marked $^{a}$ are recorded in every run record; $^{b}$ was observed at preflight and in the bundled binary; $^{c}$ was reported by the operator after the run; $^{d}$ is inferred from the unpinned package and the run date. See Appendix~\ref{app:configs}.}
\label{tab:configs}
\begin{tabularx}{\textwidth}{>{\raggedright\arraybackslash}p{3.6cm}>{\raggedright\arraybackslash}p{1.8cm}>{\raggedright\arraybackslash}X>{\raggedright\arraybackslash}p{3.2cm}>{\raggedright\arraybackslash}p{1.6cm}}
\toprule
Harness & Model & Effort, budget & Prompt & Primary metric \\
\midrule
Snyk Code CLI 1.1305.1$^{c}$ & --- & --- & none & AR recall \\
Claude Code 2.1.251$^{b}$ \secrev{} & Opus 5 & xhigh, 30 turns & benchmark system prompt + \secrev{} & AR recall \\
Claude Code 2.1.251$^{b}$ \secrev{} & Sonnet 5 & xhigh, 30 turns & as above & AR recall \\
Claude Code 2.1.251$^{b}$ \secrev{} + Snyk MCP (\texttt{snyk@latest}, 1.1307.0$^{d}$) & Opus 5 & medium, 30 turns & as above + one required \texttt{snyk\_code\_\allowbreak scan} & AR recall \\
Codex Security 0.1.24$^{a}$ (plugin 0.1.79$^{a}$) & GPT-5.6 Luna & xhigh, standard scan & built-in & AR recall \\
Codex Security 0.1.24$^{a}$ & GPT-5.6 Terra & xhigh, standard scan & built-in & AR recall \\
Codex Security 0.1.24$^{a}$ & GPT-5.6 Sol & xhigh, standard scan & built-in & AR recall \\
DeepSec 2.3.7$^{a}$ & Opus 5 & xhigh, 150 turns per batch & built-in & localized recall \\
DeepSec 2.3.7$^{a}$ & GPT-5.6 Sol & xhigh, 150 turns per batch & built-in & localized recall \\
\bottomrule
\end{tabularx}
\end{table}

\subsection{Protocol}
\label{sec:protocol}

Detection is blind: no configuration receives an advisory, a vulnerability description, a class hint or the reference set. Every run is independent. It starts in a fresh, isolated temporary workspace that contains only the application source, with no memory of earlier runs, and the answer files are never copied into it. The Claude Code runner denies reads outside the workspace; the Codex Security and DeepSec runners execute inside an operating-system sandbox whose containment is checked before any model request. Each of the 9 configurations ran once on each of the 20 applications, for 180 successful runs (the final attempt is scored; Section~\ref{sec:threats-other}), executed in early September 2026. All model traffic went through one LiteLLM gateway, so gateway latency and retries are part of the measured wall time. Turns recorded for Claude Code include sub-agent messages and can therefore exceed the 30-turn setting.

\subsection{Scoring}
\label{sec:scoring}

\paragraph{Endpoint rule.} A reported finding matches a reference vulnerability when (i) its class equals the reference class or one of its listed alternatives, and (ii) its reported locations meet the endpoint rule. Paths match after normalization, and a reported line may differ from a curated line by at most two lines. For a flow with one curated location, a match to that location is enough; for two locations, a match to either labeled endpoint is enough; for three or more, distinct reported locations must match both a labeled source and a labeled sink. Of the 62 cross-file flows, 61 fall under the strict three-or-more rule, against 29 of the 46 same-file flows. Intermediate locations never raise the bar. Matching is one-to-one: a finding can be credited to at most one vulnerability, and one reported location cannot satisfy both endpoints.

\paragraph{Why two lines.} The tolerance absorbs a tool and an annotator anchoring different lines of the same statement, such as a call that spans several lines or an assignment and the call that uses it. The headline does not depend on it. Under exact-line matching (Table~\ref{tab:exactline}), Snyk Code matches 85 instead of 89, and the ranking of the seven comparable configurations is unchanged at tolerances of zero and one line. Widening the tolerance helps the AI configurations rather than Snyk Code: at ten lines, Codex Security with GPT-5.6 Sol gains 37 matches and Snyk Code none (Section~\ref{sec:robustness}).

\begin{table}[H]
\centering
\small
\caption{Vulnerabilities matched (of 108) under the primary two-line rule and under exact-line matching.}
\label{tab:exactline}
\begin{tabular}{lrr}
\toprule
Configuration & Primary ($\pm$2 lines) & Exact line \\
\midrule
Snyk Code & 89 & 85 \\
Claude Code \secrev{}, Opus 5 xhigh & 85 & 79 \\
Claude Code \secrev{}, Opus 5 medium + Snyk MCP & 80 & 74 \\
Claude Code \secrev{}, Sonnet 5 xhigh & 73 & 55 \\
Codex Security, GPT-5.6 Luna & 45 & 15 \\
Codex Security, GPT-5.6 Terra & 35 & 6 \\
Codex Security, GPT-5.6 Sol & 34 & 2 \\
\bottomrule
\end{tabular}
\end{table}

\paragraph{DeepSec.} DeepSec exports files and lines for each finding but does not label which location is the source and which the sink. Assigning those roles ourselves would credit DeepSec with evidence it did not report, so DeepSec is scored with \emph{localized recall}: a class match plus any reported location within two lines of any curated flow location. This is a different evidence requirement. DeepSec results are always reported separately and are never ranked or combined with the seven endpoint-aware configurations.

\subsection{Metrics and statistics}
\label{sec:metrics}

The primary metric is \emph{Attacker-Reachable Vulnerability Recall}: the share of reference vulnerabilities a configuration matched under the endpoint rule. We compute it per application and average over the 20 applications (\emph{macro} recall), so that each application weighs the same regardless of how many vulnerabilities it contains, and we print the pooled count of matched vulnerabilities beside it, for example ``82.4\% (89 of 108)''. Pooled (\emph{micro}) and macro recall rank the configurations identically. Precision, the share of a configuration's findings that matched a reference vulnerability, is secondary. Because the reference set may omit real vulnerabilities, measured precision is a lower bound.

Two tools quantify uncertainty, and both are descriptive. To compare two configurations, McNemar's test looks only at the vulnerabilities that one found and the other missed, and asks how often a split at least that uneven would occur by chance if the two were equally good; we computed it for all 21 pairs of endpoint-aware configurations and report it for two of these pairs and for the supplementary control.\footnote{We report the exact McNemar test, with the less conservative mid-$p$ variant~\citep{fagerland2013mcnemar} in footnotes. Neither is adjusted for multiple comparisons or for vulnerabilities sharing an application, and we do not describe any difference as significant.} To see how much a score depends on which applications happen to be in the corpus, we redraw the 20 applications with replacement 4,000 times (seed 20260929), recompute macro recall each time, and report the middle 95\% of those values as an interval. With 20 applications of 2 to 10 vulnerabilities each, these intervals are approximate, and they say nothing about run-to-run variation, which one run per cell cannot measure.

All analysis numbers in this paper were recomputed from the public run archive by a separate Python re-implementation of the scorer, which reproduces every recorded scoring decision in all 180 runs.

\subsection{Supplementary runs}
\label{sec:supp-method}

To test two readings that the main matrix could only suggest, we ran three supplementary arms on 2026-10-01, about a month after the main runs, with the same tasks, reference set, prompts, gateway, model aliases and tool versions. Fixture trees were identical to the main runs on 17 of the 20 tasks; the other 3 differ only in files ignored by version control. Arms A and B recorded \$178.19 of known cost, with 2 attempts of unknown cost. The outcome thresholds for arms A and B, given below, were written down before any per-class or paired analysis; at that point only each arm's overall score in the harness had been seen (arm A, 76\% macro recall on 20 applications; arm B, 89\% on the 9 applications completed so far). Arm~A runs Claude Code \secrev{} with Opus~5 at medium effort without the Snyk MCP server, as an equal-effort comparison for the MCP configuration. Arm~B runs Opus~5 at xhigh effort with the benchmark's default audit prompt in place of \secrev{}; it completed on 19 of the 20 applications, because both attempts on one application were aborted externally after about four minutes, and every comparison involving it is paired on those 19 applications (105 vulnerabilities). Arm~C repeats Snyk Code twice more. For arm B, the policy effect counted as supported if removing \secrev{} added at least 5 matches among the 20 vulnerabilities in the policy classes, the discordant pairs in those classes favored the control at least 3 to 1, and recall on the other 78 vulnerabilities moved by no more than 5 percentage points. For arm A, the two configurations counted as tied if their paired difference was at most 2 vulnerabilities in either direction. The supplementary arms are reported beside the main results and are never merged into the main tables, unions or rankings.

%% file: sections/03-complementarity.tex
\section{Complementarity: SAST and AI review cover different ground}
\label{sec:complementarity}

No single configuration reaches the ceiling of what the configurations found together. Snyk Code plus one AI review does, and the two contribute different kinds of vulnerabilities.

\subsection{Main recall results}
\label{sec:headline}

\begin{figure}[H]
\centering
\includegraphics[width=\linewidth]{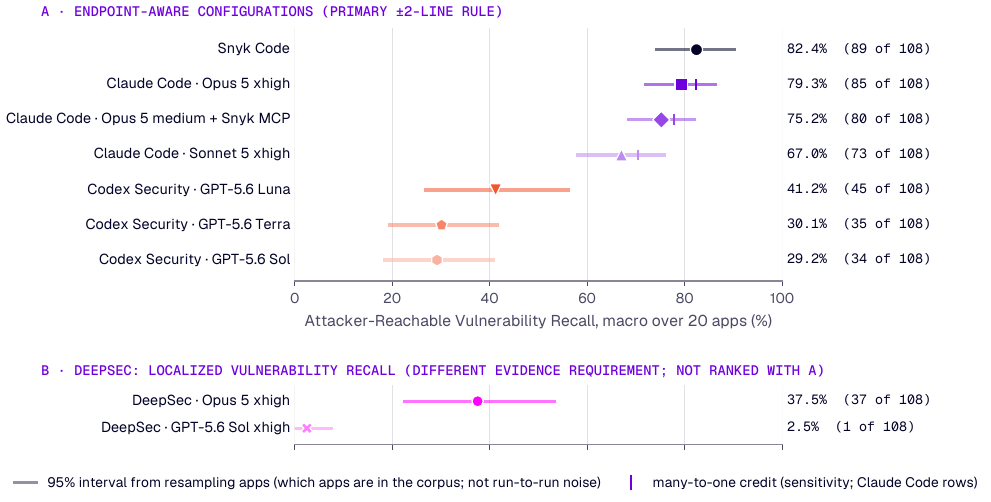}
\caption{Main recall results. (A) The seven endpoint-aware configurations: macro Attacker-Reachable Vulnerability Recall with the matched count. The horizontal bar is a 95\% interval from resampling the 20 applications; it shows how much the score depends on which applications are in the corpus, not run-to-run variation (Snyk Code's output was identical across three runs, Section~\ref{sec:robustness}). The tick on the Claude Code rows marks the score under many-to-one credit, a sensitivity check (Section~\ref{sec:robustness}); the Codex Security configurations gain at most 3 matches under it and are not marked. (B) DeepSec, scored with localized recall, on its own axis. The top two cannot be separated: they disagree on 28 vulnerabilities, 16 vs 12 (McNemar $p = 0.57$, descriptive).}
\label{fig:headline}
\end{figure}

Table~\ref{tab:headline} and Figure~\ref{fig:headline} give the main recall results against the 108 reference vulnerabilities. Snyk Code matched 82.4\% (89 of 108) and Claude Code \secrev{} with Opus~5 at xhigh effort 79.3\% (85 of 108). A corpus of this size cannot separate them: they disagree on 28 vulnerabilities, 16 matched only by Snyk Code and 12 only by Opus~5 xhigh (McNemar $p = 0.57$)\footnote{Mid-$p$ 0.46.}, and Snyk Code ranks first in 74\% of the resampled corpora. A split that close would need a corpus many times larger, with many more vulnerabilities on which the two disagree, before it could indicate a real difference; not separating them is also not evidence that they are equal. The three Codex Security configurations matched between 34 and 45 vulnerabilities; Section~\ref{sec:harness} shows that much of that gap is about which line they report, not whether they see the vulnerability. Snyk Code's precision against the reference set is lower than its recall suggests: it reported 425 findings, of which 89 matched; Section~\ref{sec:triage} shows what the rest are.

\begin{table}[H]
\centering
\small
\caption{Main recall results. All Claude Code configurations run \secrev{}. Macro recall averages the 20 per-application recalls; the matched count is pooled. The interval comes from resampling applications (Section~\ref{sec:metrics}). Precision is a lower bound because the reference set may omit real vulnerabilities. DeepSec rows use localized recall, a different evidence requirement, and are not ranked with the others.}
\label{tab:headline}
\footnotesize
\begin{tabular}{lrrrr}
\toprule
Configuration & Macro recall & Matched & 95\% interval & Precision \\
\midrule
\multicolumn{5}{l}{\emph{Attacker-Reachable Vulnerability Recall}} \\
Snyk Code & 82.4\% & 89 of 108 & 74.0--90.5 & 20.9\% (89/425) \\
Claude Code, Opus 5 xhigh & 79.3\% & 85 of 108 & 71.6--86.7 & 74.6\% (85/114) \\
Claude Code, Opus 5 medium + Snyk MCP & 75.2\% & 80 of 108 & 68.2--82.3 & 82.5\% (80/97) \\
Claude Code, Sonnet 5 xhigh & 67.0\% & 73 of 108 & 57.7--76.2 & 74.5\% (73/98) \\
Codex Security, GPT-5.6 Luna & 41.2\% & 45 of 108 & 26.6--56.5 & 20.8\% (45/216) \\
Codex Security, GPT-5.6 Terra & 30.1\% & 35 of 108 & 19.1--41.9 & 15.8\% (35/221) \\
Codex Security, GPT-5.6 Sol & 29.2\% & 34 of 108 & 18.1--41.1 & 12.7\% (34/267) \\
\midrule
\multicolumn{5}{l}{\emph{Localized Vulnerability Recall (DeepSec; not comparable)}} \\
DeepSec, Opus 5 xhigh & 37.5\% & 37 of 108 & 22.6--53.2 & 7.1\% (37/524) \\
DeepSec, GPT-5.6 Sol xhigh & 2.5\% & 1 of 108 & 0.0--7.5 & 10.0\% (1/10) \\
\bottomrule
\end{tabular}
\end{table}

\subsection{The ceiling}
\label{sec:ceiling}

\begin{figure}[H]
\centering
\includegraphics[width=\linewidth]{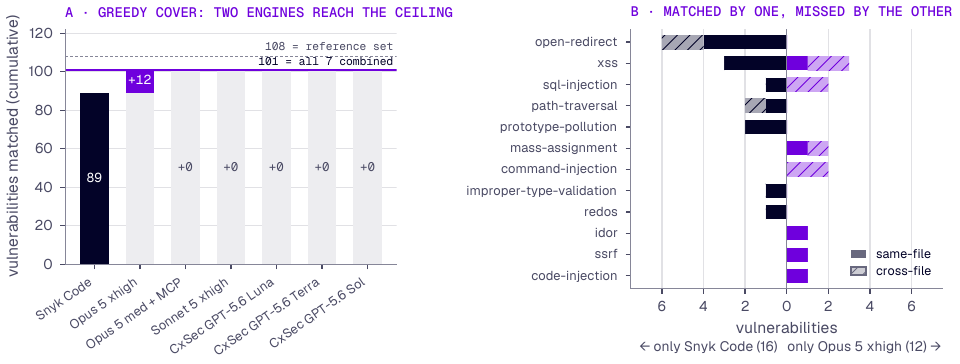}
\caption{The ceiling (endpoint-aware configurations only; DeepSec excluded). (A) Greedy cover: Snyk Code matches 89; Opus~5 xhigh adds the 12 vulnerabilities that it matched and Snyk Code did not, reaching 101, the union of all seven endpoint-aware configurations (102 after the reference-set correction); no further configuration adds any. Opus~5 xhigh also misses 16 vulnerabilities that Snyk Code matched, which the cumulative count does not show. (B) The two sets of disagreements by class: left, matched by Snyk Code and missed by Opus~5 xhigh (16); right, the reverse (12). Hatching marks cross-file flows.}
\label{fig:ceiling}
\end{figure}

\begin{figure}[H]
\centering
\includegraphics[width=\linewidth]{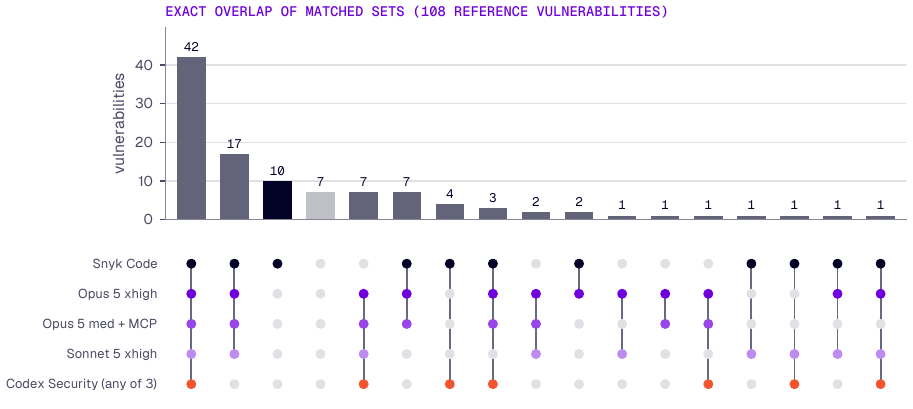}
\caption{Exact overlap of the matched sets (UpSet plot). Each bar counts the reference vulnerabilities matched by exactly the combination of configurations marked by the dots beneath it; the three Codex Security configurations are pooled. The light bar counts the 7 vulnerabilities no endpoint-aware configuration matched; the dark bar counts the 10 only Snyk Code matched (5 under many-to-one credit). DeepSec is excluded.}
\label{fig:upset}
\end{figure}

Snyk Code and Claude Code \secrev{} with Opus~5 xhigh together match 101 of the 108 vulnerabilities, exactly as many as the union of all seven endpoint-aware configurations: after those two, no other configuration adds a vulnerability (Figure~\ref{fig:ceiling}A; the full overlap is in Figure~\ref{fig:upset}). The best pair without SAST, Opus~5 xhigh with Codex Security GPT-5.6 Terra, reaches 89, the same number Snyk Code reaches alone. Snyk Code is the only one of the seven endpoint-aware configurations with unique coverage: 10 vulnerabilities that no other endpoint-aware configuration matched. They span six classes: two each of open redirect, path traversal, prototype pollution and cross-site scripting, and one each of improper type validation and ReDoS. The primary rule credits each reported finding to at most one reference vulnerability. Sometimes a single finding covers two reference vulnerabilities, for example two flows that end at the same dangerous call. Under the alternative \emph{many-to-one} credit, such a finding counts for both; the Claude Code configurations gain a few matches this way and Snyk Code none. Five of the ten vulnerabilities only Snyk Code matched are then also matched by an AI configuration, so Snyk Code's unique coverage falls to 5, and the best pair of AI configurations reaches 93, while the combined ceiling of 101 holds. With the reference-set defect of Section~\ref{sec:threats} corrected, both the pair and the union reach 102.

In a supplementary run on 19 of the 20 applications, a Claude Code configuration without the built-in review policy (Section~\ref{sec:policy}) matched more reference vulnerabilities than Snyk Code (93 vs 87 of 105). Complementarity still held: together they reached 98 of 105, Snyk Code added 5 vulnerabilities the control missed, and Snyk Code still had unique coverage against every other configuration including the control (5 vulnerabilities); the control, in turn, matched 2 that no main endpoint-aware configuration matched.\footnote{This comparison was not among the analyses whose outcome thresholds were fixed before the supplementary runs were analyzed; it is exploratory. The control is never added to the ceiling, the overlap figure or any ranking.}

\subsection{What each engine brings}
\label{sec:brings}

The two engines fill each other's gaps with different material (Figure~\ref{fig:ceiling}B). Snyk Code adds 16 vulnerabilities to Opus~5 xhigh, only 3 of them cross-file, led by open redirect (6), cross-site scripting (3), path traversal (2) and prototype pollution (2). Opus~5 xhigh adds 12 to Snyk Code, 7 of them cross-file: cross-site scripting (3), command injection (2), SQL injection (2), mass assignment (2), and one each of code injection, insecure direct object reference (IDOR) and server-side request forgery.

The authorization classes illustrate the split. On the three IDOR and mass-assignment vulnerabilities, Snyk Code matched none and the Claude Code configurations matched two or three. Three cases cannot support a class-level rate, but the pattern is what one would expect from a reasoning engine that can model which user may touch which object. In the other direction, Snyk Code matched all 11 open redirects, all 6 prototype-pollution vulnerabilities and all 16 path traversals.

\subsection{Where each engine wins, by flow shape}
\label{sec:flowshape}

\begin{figure}[H]
\centering
\includegraphics[width=\linewidth]{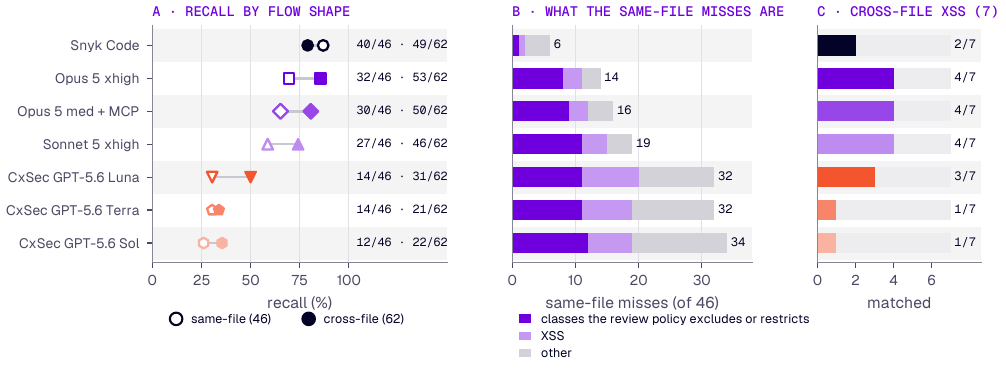}
\caption{Where each engine wins, by flow shape. (A) Recall on the 46 same-file (hollow) and 62 cross-file (filled) vulnerabilities. (B) What each configuration's same-file misses are: classes the Claude Code review policy excludes or restricts, cross-site scripting, and all other classes. (C) Cross-file cross-site scripting, the one cell where every configuration matched at most 4 of 7. The three panels share their rows: each row is the same configuration in A, B and C.}
\label{fig:flowshape}
\end{figure}

Figure~\ref{fig:flowshape} maps strengths by whether a flow stays within one file. Within a single file Snyk Code has the highest recall: it matched 40 of the 46 same-file vulnerabilities, against 32 for Opus~5 xhigh. Across files the two are close: Snyk Code matched 49 of 62 and Opus~5 xhigh 53. Claude Code's same-file misses are mostly about classes, not files: of the 14 same-file vulnerabilities Opus~5 xhigh missed, 8 are in the classes its review policy excludes or restricts (Section~\ref{sec:policy}) and 3 are cross-site scripting. For the Claude Code configurations, the blind spots are about which classes they report, not about navigating a codebase.

One cell is hard for every engine: cross-file cross-site scripting, where no endpoint-aware configuration matched more than 4 of the 7. In 5 of those 7 the data flows from server-side code to a browser page (for example, from a Go back end into a React front end); Snyk Code matched none of the 5 and Opus~5 xhigh 2. Both matched the 2 cases that stay on the client. This is the same server-to-browser boundary that recurs in Section~\ref{sec:outside}. Snyk Code's lower cross-file recall comes largely from this cell: with cross-site scripting removed, its cross-file and same-file recall are essentially equal (47 of 55 and 29 of 34).

\subsection{The cost of combining}
\label{sec:triage}

\begin{figure}[H]
\centering
\includegraphics[width=\linewidth]{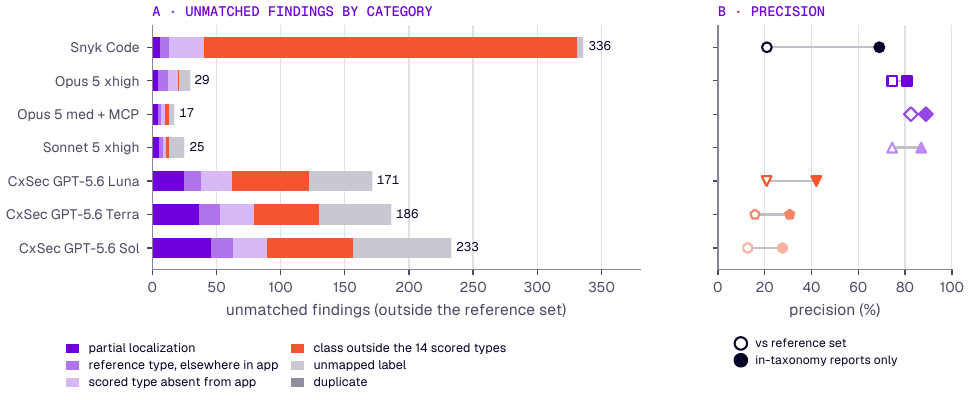}
\caption{The triage cost. (A) Findings that matched no reference vulnerability, by category; they are outside the reference set and not proven false. (B) Precision against the reference set (hollow) and over reports in the 14 scored classes only (filled); both are lower bounds. DeepSec, scored differently, is omitted: its Opus~5 backend reported 487 unmatched findings, 449 of them with labels that map to no scored class.}
\label{fig:unmatched}
\end{figure}

Combining the two engines has a triage cost. Snyk Code and Opus~5 xhigh together report 539 findings for their 101 matches. Most of the difference is Snyk Code's: 336 of its 425 findings matched no reference vulnerability, and 87\% of those are in classes outside the benchmark's 14 scored classes, led by hardcoded credentials (195) (Figure~\ref{fig:unmatched}). These may be real issues; they are outside the reference set by construction. Counting only reports in the scored classes, Snyk Code's precision is 69\% (89 of 129), against 81\% (85 of 105) for Opus~5 xhigh. Snyk Code's precision is also sensitive to how it is averaged: 54.9\% macro against 20.9\% micro, because a few applications with many findings dominate the pooled count.

\subsection{Findings outside the reference set}
\label{sec:outside}

Did the AI configurations find vulnerabilities that both the reference set and Snyk Code missed, and were they real? Of the findings outside the reference set, 23 clusters were reported by at least three configurations from at least two tool families. We reviewed the 15 of them that Snyk Code did not report and that fall in a scored class or are cross-site request forgery (CSRF). Each received a code-level first pass and an independent skeptical second pass, both performed by AI agents, and a final call by an author; exploitability was checked with AI assistance. Thirteen were real, one was borderline and one was judged to be by design; none was noise. Eight were real and new, meaning distinct from every reference vulnerability. One of the eight, a CSRF, was set aside because Snyk Code did report its root cause, which leaves seven, all in scored classes: four cross-site scripting vulnerabilities and one each of SQL injection, mass assignment and prototype pollution\footnote{One of the cross-site scripting vulnerabilities, in sassyreg, one of the authors' own demonstration applications, is described by a leftover developer comment in the fixture that names its payload. It was nonetheless absent from the reference set.} (Appendix~\ref{app:supp}, Table~\ref{tab:adjudicated}). Snyk Code did not report these flows: the only Snyk Code findings within five lines of a reported location, or of the same class in the same files, belong to three of the seven and were credited to other reference vulnerabilities. All seven were reproduced live against the running applications.\footnote{For the flask-bones cross-site scripting vulnerability, the application's 20-character limit on usernames constrains the payload: HTML injection into the administrator's session and loading of an attacker-chosen script URL were demonstrated, but full exploitation needs a script URL short enough to fit the limit.} In this run, AI configurations found at least seven real vulnerabilities in scored classes that the reference set did not contain and Snyk Code did not report. We say ``at least'' because only consensus candidates were reviewed; findings reported by a single configuration were not.

\paragraph{Why the AI configurations found them.} All seven are first-party vulnerabilities: none is a defect in a library, and the fix belongs in the application's own code in every case. Each, however, crosses at least one boundary that first-party, single-language taint analysis typically does not follow. Six pass through third-party code, such as a request validator, a SQL query builder, a framework's data binder or a user-interface library, and four cross from the server, often through stored data, into the browser's document object model. None is a plain request-to-sink flow within one language in first-party code. The AI configurations followed the data where a first-party taint model typically ends: through library semantics and across the server-to-browser hop. This is a property of these vulnerabilities and of first-party taint analysis as a category, and it matches what program-analysis work has reported: taint analyzers rely on human-written source and sink specifications~\citep{li2025iris}, and a taint analyzer for server-side request forgery in PHP builds implicit as well as explicit call graphs~\citep{ji2025artemis}. The cross-file cross-site scripting cell of Section~\ref{sec:flowshape} sits on the same server-to-browser boundary.

\paragraph{Who found them.} Codex Security configurations reported all seven, the Claude Code configurations six and DeepSec three. Codex Security's low line-level recall (Section~\ref{sec:anchoring}) therefore coexists with reporting all seven of these findings. The consensus filter, which required several configurations from at least two families, gives more weight to the family with three configurations, so these counts do not measure how broadly each family discovers issues.

These findings are reported separately and are never added to the reference set, so the main results are unchanged; nor do they support a miss rate for any engine, because they are a lower-bound count from consensus candidates rather than a sample. The final reviewer is an author, the consensus filter favors issues several tools can see, and exploitability was established by code tracing, AI-assisted checks and live reproduction rather than a full penetration test. We shared the seven findings with the Snyk Code team before publication. By 2026-10-28, one of them, the mass assignment in coffeeshop, had received a rule update deployed to production. A re-scan of the 20 applications that day, with a preview build of the Snyk CLI (1.1309.0-preview), detects that vulnerability and none of the other six; the update did not change which reference vulnerabilities Snyk Code matched (the same 89 as in the main run).\footnote{The re-scan is published separately (Data and code availability). It used the reference set as currently published, which carries the iguana correction of Section~\ref{sec:threats-other}; scored against the reference set used for all results, its matches are identical to Snyk Code's in the main run.} These post-disclosure results are reported separately and never merged into the main results, which stay as run.

%% file: sections/04-harness-policy.tex
\section{Harness policy: the harness decides what gets reported}
\label{sec:harness}

An AI security engine is a model inside a harness, and the harness's built-in instructions and pipeline visibly shape which attacker-reachable vulnerabilities reach the report. Because every reference vulnerability is attacker-reachable, a class a harness suppresses is a real miss, not noise reduction. We show three mechanisms, one per harness, and then the counterweight: inside a harness, the model still matters.

\subsection{Claude Code \secrev{}: a precision-first policy that suppresses real classes}
\label{sec:policy}

The built-in \secrev{} prompt in Claude Code 2.1.251, the version that ran, contains an explicit exclusion list and a set of precedents. Paraphrased: it excludes regular-expression denial of service (ReDoS) and other resource-exhaustion denial of service outright; it asks for prototype pollution and open redirects to be reported only at extremely high confidence; and it treats React and Angular code as safe from cross-site scripting unless unsafe rendering APIs are used. We call ReDoS, prototype pollution and open redirect the \emph{classes the built-in review policy excludes or restricts}. We chose them from the prompt text, before looking at how each configuration performed on them.

\begin{figure}[H]
\centering
\includegraphics[width=\linewidth]{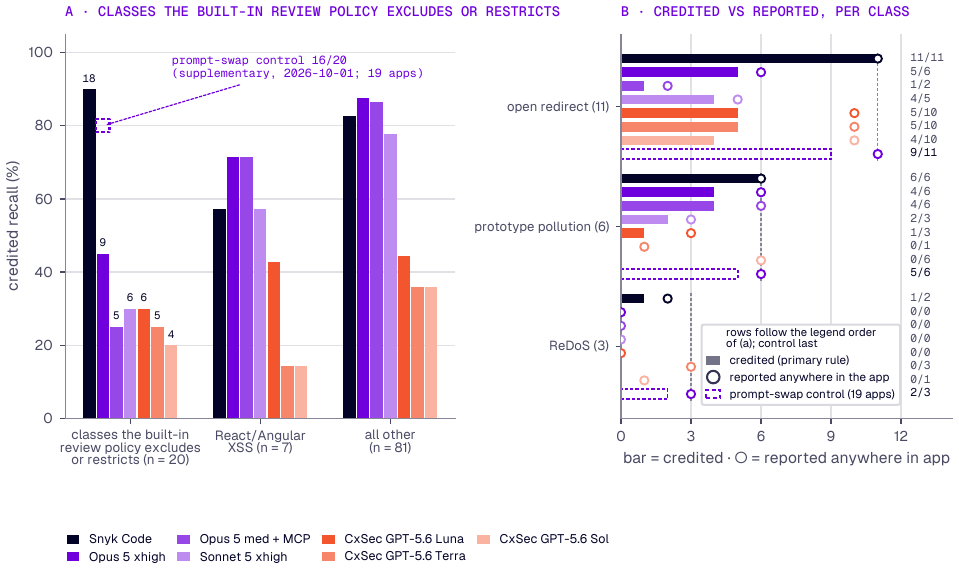}
\caption{Classes the built-in review policy excludes or restricts. The Claude Code \secrev{} prompt (version 2.1.251) excludes ReDoS and asks that prototype pollution and open redirect be reported only at extremely high confidence; these classes were chosen from the prompt text, not from the data. (A) Credited recall on the 20 vulnerabilities in those classes, the 7 cross-site scripting vulnerabilities in React and Angular applications (which do not test the corresponding precedent), and the 81 others. (B) Per class: bars show vulnerabilities credited under the two-line rule, circles show vulnerabilities for which a finding of that class was reported anywhere in the application; labels read credited/reported. The dashed outline in both panels is the supplementary prompt-swap control (Opus~5 xhigh with the benchmark's default audit prompt; 19 applications, single run, 2026-10-01), shown beside the main results and not merged into them.}
\label{fig:policy}
\end{figure}

Twenty of the 108 reference vulnerabilities fall in those classes. On them, Snyk Code matched 18 and Claude Code \secrev{} with Opus~5 xhigh 9. On the 81 vulnerabilities in other classes (excluding seven cross-site scripting vulnerabilities in React and Angular applications, discussed below), the two are level: Opus~5 xhigh matched 71 and Snyk Code 67 (they disagree on 16, split 10 to 6). The policy classes account for more than the whole headline gap between them (Figure~\ref{fig:policy}A).

The per-class evidence is uneven, and we state where it is weak (Figure~\ref{fig:policy}B).
\begin{itemize}
  \item \emph{ReDoS.} No Claude Code configuration reported a single ReDoS finding anywhere in the three applications that contain one, consistent with the outright exclusion.
  \item \emph{Open redirect.} There are 11 open-redirect vulnerabilities in the reference set. Counting a configuration as having \emph{reported} one when it produced at least one open-redirect finding anywhere in the same application, whether or not at the right location: the Claude Code configurations reported 2 to 6 of the 11; each of the three Codex Security configurations, whose harness carries no such instruction, reported 10 of the 11; and Snyk Code reported and was credited for all 11.
  \item \emph{Prototype pollution.} Here the evidence is weaker. Opus~5 at both effort settings reported all 6 and was credited for 4, so this class is not suppressed for Opus.
\end{itemize}
When Opus~5 xhigh missed a vulnerability in these classes, it usually reported no finding of that class at all (8 of its 11 misses), which is what suppression looks like and also what non-detection looks like.

Codex Security's contrast shows what a harness without the policy surfaces, not that it matches more. Its three configurations reported a finding of the right class for 13 to 17 of the 20, but were credited for few of them, because of how they localize (Section~\ref{sec:anchoring}): 4 or 5 of the 11 open redirects and none of the ReDoS cases. The correct reading is that Codex Security's harness surfaces these classes far more often than Claude Code's: its configurations reported a finding of the right class for 13 to 17 of the 20, the Claude Code configurations for 8 to 12. The seven React and Angular cross-site scripting vulnerabilities do not test the React and Angular precedent: at least four of them use unsafe rendering APIs that the precedent does not exempt, and the Claude Code configurations matched 4 or 5 of the 7.

\paragraph{A prompt-swap control.} The pattern above is consistent with the policy shaping the result, but the main matrix cannot separate policy from model. We therefore ran a supplementary control on 2026-10-01: Opus~5 at the same xhigh effort, with the benchmark's default audit prompt in place of \secrev{}, on 19 of the 20 applications (one application failed twice and is excluded; it contains none of the 20 policy-class vulnerabilities). The outcome thresholds were fixed before the per-class analysis (Section~\ref{sec:supp-method}), and the result met the threshold for ``supported''. With the built-in review policy removed, the same model at the same effort matched 16 of the 20 vulnerabilities in the classes the policy excludes or restricts, against 9 with \secrev{}, with little change elsewhere: 71 against 69 of the 78 other non-React/Angular vulnerabilities on those applications, a difference of 2.6 percentage points. All 7 vulnerabilities on which the two disagreed in the policy classes favored the control (7 vs 0; McNemar $p = 0.016$, descriptive)\footnote{Mid-$p$ 0.008.}, and the control reported a finding of the right class for all 20, against 12 with \secrev{}. Per class, it matched 9 of the 11 open redirects (against 5), 5 of the 6 prototype-pollution vulnerabilities (against 4) and 2 of the 3 ReDoS cases (against none). This is consistent with the policy, not the model, accounting for most of that gap. It rests on a single run per arm, about a month after the main runs; the gateway, model aliases, tools, tasks and reference set were identical, but changes on the model provider's side cannot be excluded.

\subsection{Codex Security: the right function, a line above the dangerous call}
\label{sec:anchoring}

\begin{figure}[H]
\centering
\includegraphics[width=\linewidth]{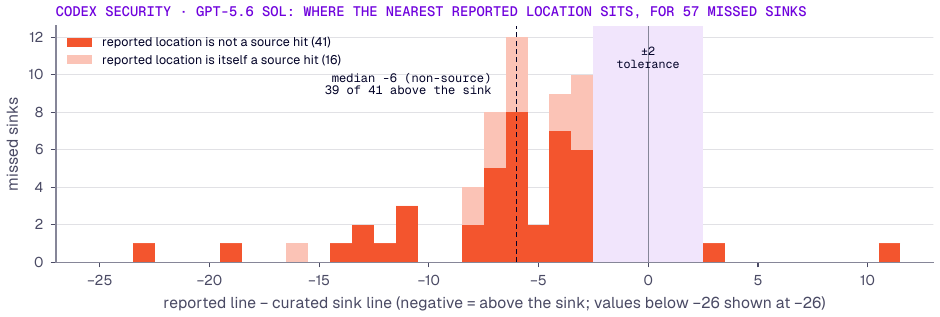}
\caption{Codex Security with GPT-5.6 Sol: for each of the 57 missed curated sinks with a reported location in the same file, the offset of the nearest reported location from the sink line (negative means above). Lighter bars are reported locations that are themselves a curated source, for which ``above'' is expected; among the other 41, 39 lie above the sink, median six lines. The shaded band is the two-line tolerance.}
\label{fig:offsets}
\end{figure}

Codex Security often sees the vulnerability and reports a location just above it. With GPT-5.6 Sol it located 80 of the 108 vulnerabilities at file level, but 34 under the two-line rule; matched by class alone, it would reach 88, which is an upper bound because it ignores location. When it missed a curated sink, its nearest reported location in the same file was, in 53 of the 57 cases, one it had itself labeled a sink, and that location sat above the curated line in 39 of 41 cases in which the reported location was not also a source, a median of six lines above (Figure~\ref{fig:offsets}). Across the three Codex Security configurations, the line reported is often the enclosing function or route header: by a simple syntactic heuristic, 17 of 38 distinct reported locations in a set of verified examples are function, route or method headers.\footnote{Examples are drawn only from the 10 fixtures whose checkout reproduces the recorded run fingerprint. The heuristic is a regular expression over the reported line and is illustrative only.} Listing~\ref{lst:conduit} shows a typical case.

\par\noindent\begin{minipage}{\linewidth}
\begin{lstlisting}[language=Python, caption={A command injection in the conduit application (\texttt{conduit/articles/pdf.py}). Codex Security with GPT-5.6 Sol reported line 10, the function header; the curated sink is line 16, six lines below.}, label={lst:conduit}, numbers=left, firstnumber=10]
def render_pdf(body, label):
    if not os.path.isdir(EXPORTS_DIR):
        os.makedirs(EXPORTS_DIR)
    command = 'pandoc --from=markdown --to=pdf --output="{0}/{1}.pdf" -'.format(
        EXPORTS_DIR, label,
    )
    process = subprocess.Popen(command, shell=True, stdin=subprocess.PIPE)
\end{lstlisting}
\end{minipage}

\begin{figure}[H]
\centering
\includegraphics[width=\linewidth]{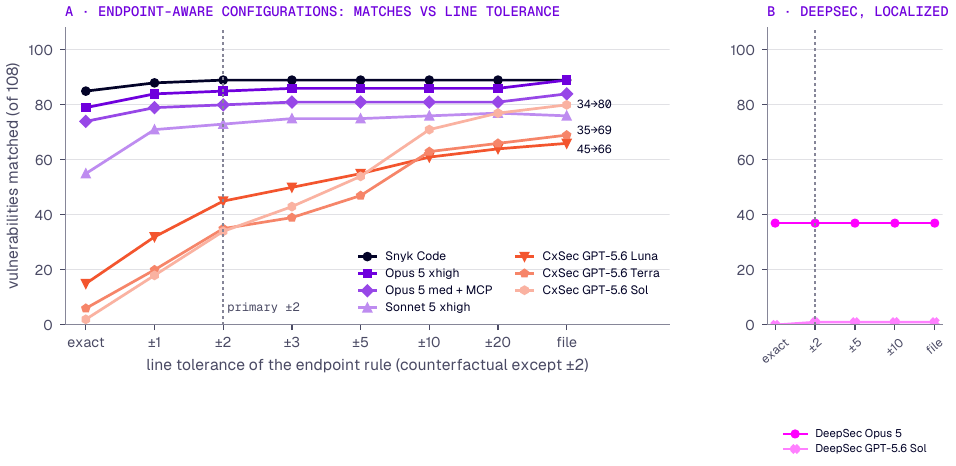}
\caption{Matches against line tolerance. (A) The seven endpoint-aware configurations under the endpoint rule, from exact line to file level; only the two-line setting is used for the results, the others are counterfactual rescorings of the same findings. Labels give the change from the two-line setting to file level for Codex Security. (B) DeepSec under its own localized rule, on a separate axis.}
\label{fig:tolerance}
\end{figure}

A harness can detect a vulnerability and still hand the developer the wrong line. Line-level evidence is what makes a finding actionable, and it is what the endpoint rule measures. The tolerance curve (Figure~\ref{fig:tolerance}) shows how much of Codex Security's score depends on that choice: its matches rise steeply as the tolerance widens, while Snyk Code and the Claude Code configurations barely move.

\subsection{DeepSec: a recall ceiling fixed before the model runs}
\label{sec:deepsec}

\begin{figure}[H]
\centering
\includegraphics[width=\linewidth]{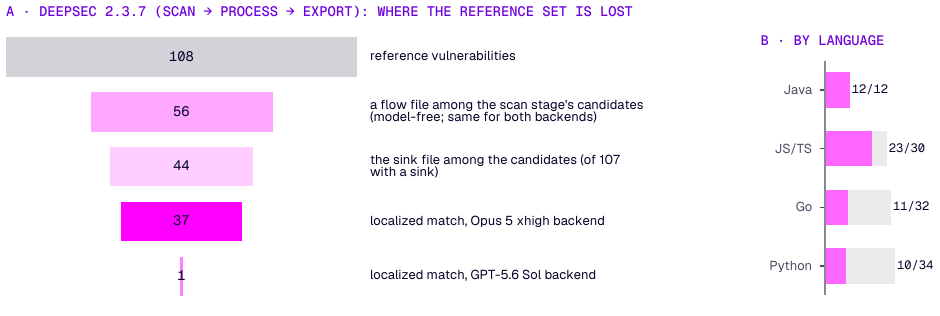}
\caption{Where DeepSec~2.3.7 loses the reference set. (A) Of 108 reference vulnerabilities, 56 had a flow file among the scan stage's candidates (model-free, identical for both backends) and 44 had their sink file there; localized matches follow for each backend. (B) Candidate coverage by language family. Localized recall is DeepSec's primary metric and is not comparable with Attacker-Reachable Vulnerability Recall.}
\label{fig:funnel}
\end{figure}

DeepSec runs a model-free scan stage that decides which files the model will see. That stage queued at least one file from the curated flow for only 56 of the 108 vulnerabilities, and the sink file for 44 of the 107 vulnerabilities that have a labeled sink (Figure~\ref{fig:funnel}). Coverage differed sharply by language: all 12 Java vulnerabilities and 23 of 30 in JavaScript and TypeScript, but 11 of 32 in Go and 10 of 34 in Python. In three applications the stage produced two or fewer candidate files; in conduit the only candidate was the Dockerfile. The candidate set was identical for both model backends on every application, and neither backend matched a vulnerability whose files were not selected. Conditional on selection, the Opus~5 backend's localized recall was 37 of 56 and the GPT-5.6 Sol backend's 1 of 56. The GPT-5.6 Sol backend exported no findings in 16 of its 20 runs, although every run satisfied the harness's success contract; its median run recorded 4 turns against 116 for the Opus~5 backend, and 9 of its runs recorded 3 or fewer. The records do not show whether this reflects the backend's behavior inside DeepSec or an integration problem, so this row describes the configuration as executed and says nothing about the model. We report DeepSec separately, under localized recall, and attribute these results to the DeepSec~2.3.7 configuration as run.

\subsection{The counterweight: models still matter inside a harness}
\label{sec:models-matter}

None of this means the model is irrelevant. Inside the same harness, at the same effort and with the same prompt, Claude Code with Opus~5 matched 14 vulnerabilities that Sonnet~5 missed, against 2 the other way (McNemar $p = 0.004$, descriptive),\footnote{Mid-$p$ 0.002.} a macro difference of 12.2 percentage points (95\% interval from resampling applications: 4.7 to 19.7). Inside DeepSec, the two backends faced the identical candidate set and reached 37 and 1 of its 56 vulnerabilities. Our harness claim rests on identified mechanisms (the policy text and its control, the anchoring offset, the pre-model file selection), not on aggregate score gaps between harnesses, which also differ in model.

The policy also has a cost beyond the classes it excludes or restricts. On the same 19 applications, the prompt-swap control matched 93 of 105 vulnerabilities against 83 for \secrev{} (12 vs 2 discordant), at 0.63 times the cost, 0.52 times the wall time and 0.84 times the tokens, with similar precision against the reference set (73.2\% against 74.8\%). This is a single supplementary run, and it is not merged into the main results; it suggests that the built-in policy is a choice with consequences, and that a team adopting a productized harness inherits it.

%% file: sections/05-efficiency.tex
\section{Efficiency: what coverage costs}
\label{sec:efficiency}

Coverage and cost were only loosely related in this study: the most expensive configurations were not the most effective, and the deterministic scan was more than an order of magnitude faster than any AI configuration (8 minutes in total against at least 2.4 hours). We report wall time, dollars and tokens, and we compare like with like: dollars only where a cost was recorded, and tokens only within a provider family.

\paragraph{What is counted.} Cost is the known cost recorded by the harness for each run, \$1{,}594.08 in total across the main matrix. Snyk Code has no per-scan model cost, so it appears in the time comparisons and is marked separately in the cost comparisons; its license cost is outside this study. DeepSec with the GPT-5.6 Sol backend recorded no dollar cost, which is unavailable, not zero. Every cost, token and time figure comes from a single run per configuration and application. Agent token use can vary widely between runs of the same task, by up to 30 times in total tokens on SWE-bench Verified~\citep{bai2026tokens}, so ratios such as ``35 times'' below are single observations, not estimates with a known spread.

\subsection{Time}

Snyk Code scanned all 20 applications in 8 minutes. The comparable AI configurations took between 2.4 hours (Claude Code \secrev{} with Opus~5 at medium effort and the Snyk MCP server) and 8.9 hours (Codex Security with GPT-5.6 Sol); DeepSec with the Opus~5 backend took 17.8 hours. Per matched vulnerability, that is 5 seconds for Snyk Code and between 107 and 942 seconds for the endpoint-aware AI configurations (Figure~\ref{fig:permatched}A). Snyk Code alone forms the recall-versus-time frontier (Figure~\ref{fig:costtime}B). Wall time includes the harnesses' multi-stage pipelines and the shared gateway, so it measures each configuration as deployed, not model speed alone.

\subsection{Dollars and tokens}

\begin{figure}[H]
\centering
\includegraphics[width=\linewidth]{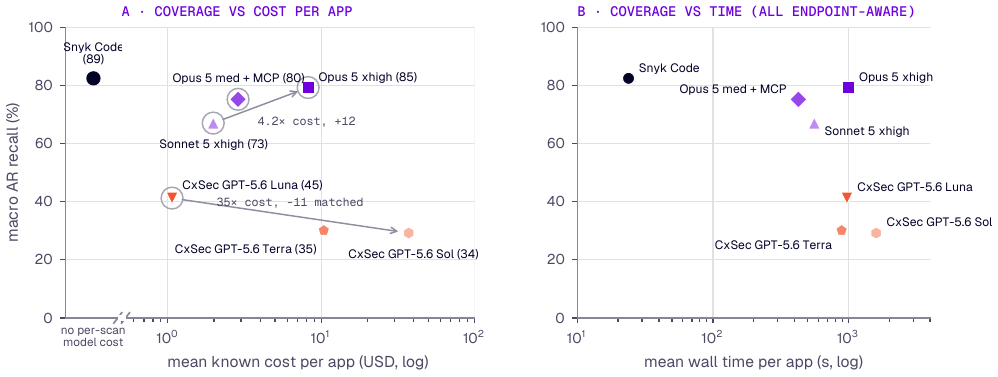}
\caption{Coverage against cost and time (endpoint-aware configurations; DeepSec is scored differently and excluded). (A) Macro recall against the mean known cost per application, on a log scale; Snyk Code has no per-scan model cost and is shown in a separate column, not on the dollar axis. Ringed points form the recall-versus-cost frontier; arrows mark the two same-harness contrasts discussed in the text. (B) Macro recall against mean wall time per application, on a log scale. One run per configuration and application; every ratio is a single observation.}
\label{fig:costtime}
\end{figure}

\begin{figure}[H]
\centering
\includegraphics[width=\linewidth]{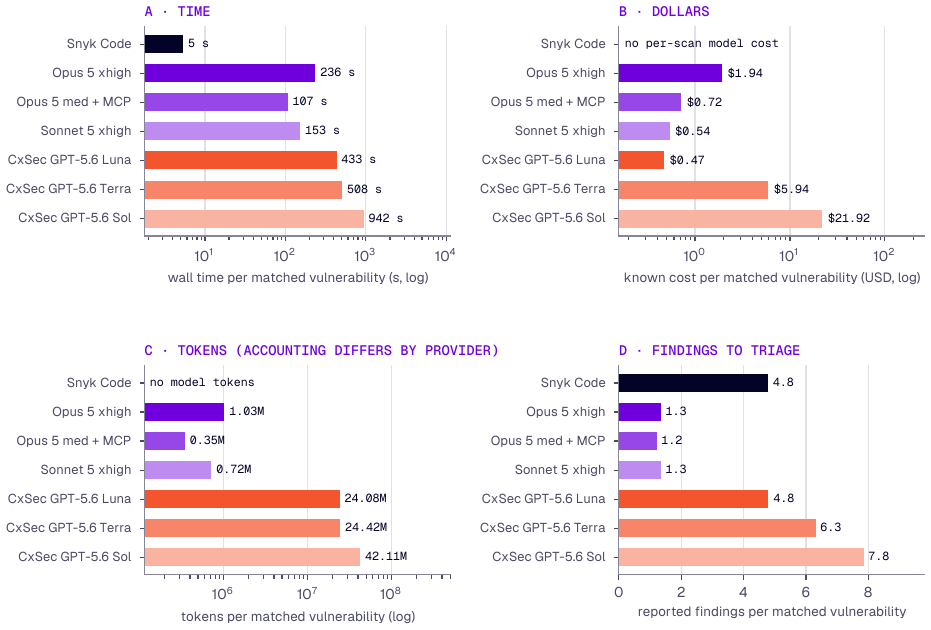}
\caption{Resources per matched reference vulnerability: totals over the 20 applications divided by the number of vulnerabilities matched. Rows are in the same order in every panel. (A) Wall time, log scale. (B) Known dollar cost, log scale; Snyk Code has no per-scan model cost. (C) Tokens, log scale; provider accounting differs, so compare within a family. (D) Reported findings per matched vulnerability, a measure of review volume.}
\label{fig:permatched}
\end{figure}

Among the configurations with a known cost, the cheapest per matched vulnerability were Codex Security with GPT-5.6 Luna (\$0.47), Claude Code with Sonnet~5 at xhigh effort (\$0.54) and Claude Code with Opus~5 at medium effort plus the Snyk MCP server (\$0.72); the most expensive was Codex Security with GPT-5.6 Sol (\$21.92) (Figure~\ref{fig:permatched}B). The recall-versus-cost frontier consists of GPT-5.6 Luna, Sonnet~5 xhigh, Opus~5 medium + MCP and Opus~5 xhigh (Figure~\ref{fig:costtime}A). Measured in tokens, the Claude Code configurations used 0.35 to 1.03 million tokens per matched vulnerability and the Codex Security configurations 24 to 42 million (Figure~\ref{fig:permatched}C). The two providers account for cached input differently, so the token figures should be compared within a family, or read only as an order of magnitude across families.

A last resource is the reviewer's time. Per matched vulnerability, the Claude Code configurations reported 1.21 to 1.34 findings, Snyk Code 4.78 and the Codex Security configurations 4.80 to 7.85 (Figure~\ref{fig:permatched}D). Findings outside the reference set are not necessarily false (Section~\ref{sec:outside}), so this measures how much a team would have to review, not how much of it is noise.

\subsection{More spend did not reliably buy more coverage}

Two contrasts inside a single harness show that the relationship between spend and coverage depends on the configuration.
\begin{itemize}
  \item \emph{A more expensive model that matched less.} Within Codex Security, all at xhigh effort, GPT-5.6 Sol cost 35.0 times as much as GPT-5.6 Luna and matched 11 fewer vulnerabilities. This is a comparison of model tiers at equal effort, not of reasoning effort.
  \item \emph{A more expensive model that matched more.} Within Claude Code \secrev{}, at the same xhigh effort and with the same prompt, Opus~5 cost 4.2 times as much as Sonnet~5 and matched 12 more. Here the spend did buy coverage.
\end{itemize}
Reasoning effort can be compared with harness and model held fixed only once, for Claude Code \secrev{} with Opus~5, across runs about a month apart: at xhigh effort it matched 85 vulnerabilities, and at medium effort, in the supplementary equal-effort arm described below, 81. Frameworks that weigh accuracy against cost~\citep{erol2025costofpass,fan2025sweeffi} would count the GPT-5.6 Sol configuration as dominated: it costs more and matches less than its own harness's cheaper option.

\subsection{The Snyk MCP configuration}

Claude Code \secrev{} with Opus~5 at medium effort and one required Snyk Code scan through the MCP server matched 80 vulnerabilities, every one of which Opus~5 at xhigh effort also matched. It did so at 0.35 times the cost, 0.43 times the wall time and 0.32 times the tokens of the xhigh configuration, with fewer findings outside the reference set (17 against 29). Effort, MCP access and the appended prompt instruction all differ between the two, so this is a cheaper configuration that kept 80 of the 85, not an effect of grounding.

The supplementary equal-effort arm separates effort from MCP access. Run at the same medium effort without the MCP server, Claude Code \secrev{} with Opus~5 matched 81 vulnerabilities against the MCP configuration's 80: they disagreed on 5, split 2 to 3, and their macro recall was 75.6\% against 75.2\%, a tie by the threshold fixed before the paired analysis. This is consistent with effort, rather than MCP access, accounting for the gap to Opus~5 xhigh. The MCP configuration's run was cheaper and quieter (\$57.44 against \$79.60, 2.4 against 2.8 hours, 27.8 against 46.6 million tokens, 17 against 24 findings outside the reference set), but these are two single runs about a month apart, and we report the difference as observed, not as an effect of grounding (Appendix~\ref{app:supp}, Figure~\ref{fig:equaleffort}).

\subsection{Scaling with application size}

Wall time tracked application size most closely for Codex Security with GPT-5.6 Terra (Spearman $\rho = +0.78$) and the two DeepSec configurations ($+0.68$ and $+0.71$), moderately for GPT-5.6 Luna ($+0.57$), and weakly for GPT-5.6 Sol ($+0.39$), Snyk Code ($+0.31$) and the Claude Code configurations ($+0.07$ to $+0.36$). Appendix~\ref{app:supp} (Figure~\ref{fig:scaling}) plots the endpoint-aware configurations. Application sizes range from about 480 to 60,000 non-blank source lines; they were counted approximately, and 20 applications are too few for more than a descriptive reading.

%% file: sections/06-robustness.tex
\section{Robustness of the results}
\label{sec:robustness}

Benchmark rankings can change with scoring details that look minor. Perturbations such as reordering answer choices or changing how an answer is selected move language models by up to eight places on multiple-choice leaderboards~\citep{alzahrani2024benchmarks}; in a security-scanner benchmark, the leader changes between the F1 and F3 scores~\citep{pellew2026realvuln}; and in conversational-memory retrieval, changing only which stored item receives credit can flip conclusions about memory architectures~\citep{panthi2026scoringtargets}. We therefore rescored the saved outputs of the main matrix under 20 alternative scoring rules, without rerunning any configuration, and asked whether the two lead findings depend on the rule we chose. They do not; the middle of the table does.

\begin{figure}[H]
\centering
\includegraphics[width=0.82\linewidth]{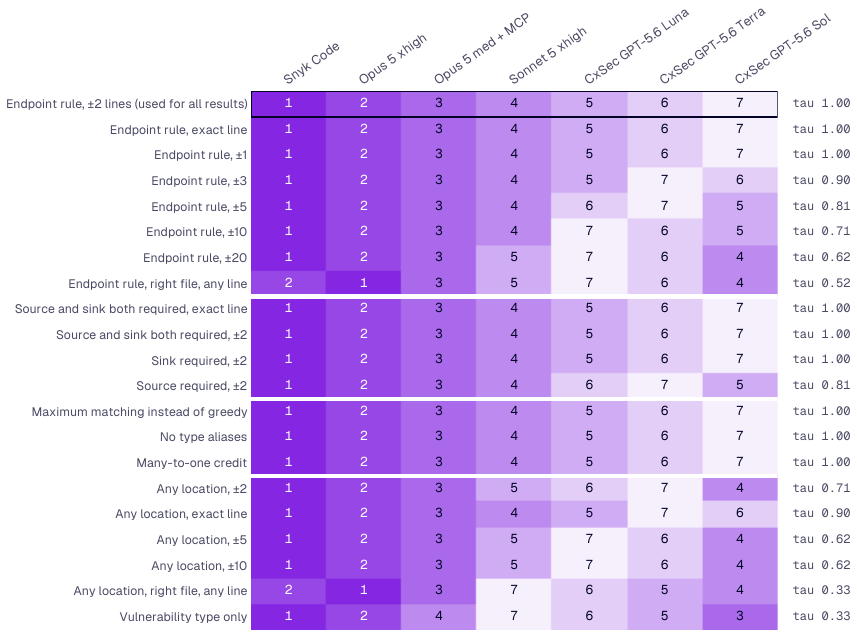}
\caption{Rank of each endpoint-aware configuration by macro recall under the scoring rule used for all results (outlined row) and 20 alternative rules, grouped as line tolerance, endpoint roles, matching and credit, and location-only or type-only rules. The right column gives Kendall's $\tau$ against the outlined row. All alternatives rescore the same saved outputs; DeepSec is excluded because it reports no endpoint roles.}
\label{fig:ranks}
\end{figure}

\paragraph{The top two are stable.} Snyk Code and Claude Code \secrev{} with Opus~5 xhigh are the two highest-ranked endpoint-aware configurations under every one of the 20 alternatives (Figure~\ref{fig:ranks}). Snyk Code ranks first under 18. The two exceptions are the most lenient location rules, which credit any reported line in the right file; under both, Opus~5 xhigh ranks first. This agrees with Section~\ref{sec:headline}: the two are not separable on these 108 vulnerabilities, and their order should not be read as a result.

\paragraph{The middle reorders.} Rank agreement with the rule used for all results ranges from a Kendall $\tau$ of 1.00 (exact line, $\pm 1$ line, both endpoints required, sink required, maximum matching instead of greedy, no type aliases, many-to-one credit) down to 0.33 (type only, and any location in the right file). The reordering is concentrated among the Codex Security configurations and Sonnet~5 xhigh. The top four change membership under 7 of the 20 alternatives, and in each case Codex Security with GPT-5.6 Sol replaces Sonnet~5 xhigh. Line tolerance is the main driver, as Section~\ref{sec:anchoring} showed: widening it from $\pm 2$ to $\pm 10$ lines adds 37 matches for GPT-5.6 Sol, 28 for Terra and 16 for Luna, at most 3 for any Claude Code configuration, and none for Snyk Code (Figure~\ref{fig:tolerance}).

\paragraph{Aggregation, matching and single applications.} Macro and micro recall give the same order for all seven endpoint-aware configurations. Replacing greedy matching with maximum matching changes no configuration's count. Removing the reference set's alternative class names changes one match (GPT-5.6 Sol, 34 to 33). Under many-to-one credit (Section~\ref{sec:ceiling}), Opus~5 xhigh rises from 85 to 89 matches, equal to Snyk Code's unchanged 89, and the macro order does not change (Appendix~\ref{app:supp}, Figure~\ref{fig:credit}). Leaving out one application at a time changes some rank in 4 of the 20 cases and the top rank in none.

\paragraph{Repeatability of the deterministic scan.} In two supplementary repetitions run a month after the main matrix, Snyk Code returned byte-identical findings on all 20 applications, and the same 89 matches each time.\footnote{Snyk CLI 1.1305.1, as reported by the operator. The AI configurations ran once per application, and we make no claim about their run-to-run variation.}

The sample is small. With 108 reference vulnerabilities in 20 applications, a study of this size can establish large differences, but not the order of closely matched configurations~\citep{card2020littlepower}. The rescoring above shows which conclusions survive changes in the scoring rule; it does not add statistical power.

%% file: sections/07-threats.tex
\section{Threats to validity}
\label{sec:threats}

\subsection{Reference-set construction, independence and authorship}
\label{sec:threats-gt}

\paragraph{How Snyk Code was involved.} Snyk Code was used during seeding and curation only as a reference for the location format and to standardize vulnerability-class names on one vocabulary (for example, \texttt{javascript/DOMXSS} becomes \texttt{xss}). The seeding prompts instructed that vulnerabilities Snyk Code did not detect be preserved, and that realism not be weakened to raise its recall. This is a property of the instructions given; we cannot verify independently that every seeded vulnerability followed it. Five of the applications had previously been used in Snyk product demonstrations: the four written by Snyk employees (Section~\ref{sec:corpus}) and the third-party application halloween. None was used in Snyk Code development or rule testing.

\paragraph{Possible biases, in both directions.} Three features of the construction could tilt the comparison, and they point in opposite directions.
\begin{itemize}
  \item \emph{Seeding models and prompts.} GPT-5.6 Luna, GPT-5.6 Terra, GPT-5.6 Sol and Anthropic models helped seed vulnerabilities that the same model families were later asked to find. A model may recognize patterns like its own, which would tend to favor the AI configurations. Without a record of which model seeded which application, we cannot test this. The seeding prompts were also not standardized: they varied from session to session, and some included examples of vulnerable code patterns from a Snyk vulnerability database. Vulnerabilities modeled on such patterns could be easier for Snyk Code to recognize, which would tend to favor it; which seeded vulnerabilities came from such prompts was not recorded. Codex Security, which ran exactly those GPT-5.6 models, nonetheless scored lowest of the seven endpoint-aware configurations, and most of its losses are about location rather than detection: matched by class alone, its three configurations reach 79 to 88 vulnerabilities.
  \item \emph{Class names follow Snyk Code's vocabulary.} Findings labeled outside that vocabulary could go uncredited. Labels that map to no scored class are rare for Snyk Code (5 of 425 findings) and more common for Codex Security (49 to 77 per configuration). The measured effect on matching is small: removing the reference set's alternative class names changes one match, and for two reference vulnerabilities that the adjudication found to be exploitable as cross-site scripting but not labeled so, adding that name raises the AI configurations by at most one match each and leaves Snyk Code unchanged.
  \item \emph{Prior demonstration use.} The five applications used in Snyk demonstrations could favor Snyk Code. Without them the main result holds: Snyk Code matches 74 and Opus~5 xhigh 72 of the remaining 87 vulnerabilities, they remain the top two, and their union still equals the union of all seven endpoint-aware configurations (83); on the five alone, Snyk Code matches 15 and Opus~5 xhigh 13 of 21.
\end{itemize}
The directions differ, and the net effect is unknown.

\paragraph{Authorship.} All authors are employees of Snyk, which develops Snyk Code. The Snyk Code product owner is a co-author and contributed to the review of this paper. After seeing the seven findings of Section~\ref{sec:outside}, his team shipped a rule update for one of them; Section~\ref{sec:outside} gives the post-disclosure status. All main results in this paper are reported as run, before that update.

\subsection{Remaining threats}
\label{sec:threats-other}

\paragraph{One run per configuration and application.} Each of the 180 cells ran once, so run-to-run variation of the AI configurations is unmeasured, and every difference between them is a single observation. Five failed attempts preceded the scored success in 4 of the 180 runs; three of them had been marked as not retryable by the harness and were run again anyway. Only the final attempt is scored, which is a mild form of re-running until success. Snyk Code's repeatability is shown separately (Section~\ref{sec:robustness}).

\paragraph{A defect in the reference set.} One reference vulnerability, a command injection in iguana, cannot be matched under the endpoint rule. Its curated flow lists one location twice, first without a role and then as the sink; when duplicate locations are merged, the scorer keeps the first occurrence and the sink label is lost. Correcting the entry raises Claude Code \secrev{} with Opus~5 at medium effort plus the Snyk MCP server from 80 to 81 and Opus~5 xhigh from 85 to 86, raises the union of all endpoint-aware configurations from 101 to 102, and changes no other configuration. The harness now carries the fix, but every result in this paper uses the reference set as frozen for the runs.

\paragraph{Completeness of the reference set.} The reference set can omit real vulnerabilities; Section~\ref{sec:outside} confirms at least seven. Adding omissions would raise the denominator for every configuration and the numerator only for those that reported them, so measured recall can move in either direction, while measured precision can only rise. Precision is therefore a lower bound with respect to omissions. By design the reference set also covers 14 vulnerability classes; reports in other classes, such as hard-coded credentials, are outside it whether or not they are real, and most of Snyk Code's findings outside the reference set fall there (Section~\ref{sec:triage}).

\paragraph{Scoring choices.} The two-line tolerance, the endpoint rule and one-to-one credit are choices, not facts about the tools. Section~\ref{sec:robustness} shows which conclusions depend on them. The alternative rules were computed by our own port of the scorer. It reproduces every recorded scoring decision of the harness, and one alternative path was checked against harness output; the others reuse the same validated components but have no harness counterpart to check against.

\paragraph{Configurations, not models.} Each result belongs to a full configuration: harness, version, mode, model and effort. Codex Security ran in its standard full-repository mode, not its deep mode, and its scans reported partial coverage on 7, 6 and 3 of the 20 applications for GPT-5.6 Luna, Terra and Sol; partial scans are scored as returned. DeepSec results belong to its scan, process and export pipeline. The design cannot separate the contribution of the harness from that of the model. Likewise, Snyk Code is the only SAST engine evaluated, and it is the authors' own product: statements about deterministic SAST in this paper are measured on Snyk Code and may not hold for other engines.

\paragraph{Execution environment.} All model traffic went through one LiteLLM gateway, so its latency and failures are part of wall time and retry history. The main matrix was executed in two phases: 140 runs at one harness commit, and the 40 DeepSec runs at a later commit. DeepSec first ran with 30 turns per batch; after 10 successful runs and one that failed on the turn limit, that profile was stopped and replaced by one with 150 turns, DeepSec~2.3.7's default, and all 40 DeepSec runs were executed fresh. The 11 earlier records remain in the parent archive and are not used. The later execution imported the earlier runs only after checking that their configuration fingerprints matched, and the imported records are byte-for-byte identical to the originals.

\paragraph{Version recording.} Codex Security (0.1.24, plugin 0.1.79) and DeepSec (2.3.7) versions are recorded on every run. The Claude Code version (2.1.251) is known from the preflight check and the binary bundled with the pinned Agent SDK, not from per-run records. The archive does not record the Snyk CLI version. The operator reported the host CLI behind the Snyk Code rows as 1.1305.1 after the run; that version was published before the runs. The Snyk MCP server was launched unpinned (\texttt{snyk@latest}); by publish dates it would have resolved to 1.1307.0 on the run date, which is inferred, not observed. Exact re-execution of the Snyk Code and MCP rows is therefore bounded; rescoring is not affected. Our check reproduces the source fingerprint recorded at run time for only 10 of the 20 current fixture checkouts (the check approximates the harness's file ordering), so the application sizes and the code examples we give come from checkouts that may differ slightly from what the configurations saw. The run records do not store the agent's stop reason, so we cannot say whether any Claude Code run ended at the 30-turn limit.

\paragraph{Cost.} The \$1{,}594.08 is known cost. It excludes 44 attempts whose cost is unavailable, among them every Snyk Code run (no model cost) and every DeepSec run with the GPT-5.6 Sol backend, and \$175.13 spent on an earlier, discarded DeepSec profile. No unavailable cost is treated as zero.

\paragraph{Statistics.} Intervals come from resampling 20 applications of 2 to 10 vulnerabilities each, so they are approximate. The McNemar tests are descriptive: unadjusted for multiple comparisons and blind to clustering within applications. For one pair of the 21 endpoint-aware pairs, the exact test and the mid-$p$ fall on different sides of 0.05: the two Opus~5 configurations, which disagree on 5 vulnerabilities, all in Opus~5 xhigh's favor.

\paragraph{Corpus scope.} Several applications are intentionally vulnerable demonstration projects, and many of the reference vulnerabilities were seeded. Java is thin (2 applications, 12 vulnerabilities). How these results carry over to production code bases, other languages or larger applications is untested.

%% file: sections/08-related-work.tex
\section{Related work}
\label{sec:related}

\paragraph{Static analysis, language models and their combination.} Combining static analyzers has long been known to find more than any one of them. On 192 real vulnerabilities in 27 C projects, single analyzers missed between 47\% and 80\%, and combining them reduced the misses to between 30\% and 69\%, at the cost of 15 percentage points more functions flagged~\citep{lipp2022staticc}; studies of static bug detectors and of Java SAST tools ask the same question~\citep{habib2018howmany,li2023javasast}. Comparisons of SAST tools with language models report that the models detect more vulnerabilities but with more false positives, and that combining the two families mitigates the weaknesses of each~\citep{zhou2024sastllm}; another finds that language models misplace issues at line granularity and recommends a hybrid pipeline~\citep{gnieciak2025llmsast}. CASTLE compares 13 static analyzers, 10 language models and 2 formal verifiers on 250 small programs and finds that the models' accuracy falls as code grows~\citep{dubniczky2025castle}. RealVuln ranks rule-based, general-purpose and security-specialized scanners on 26 intentionally vulnerable Python repositories~\citep{pellew2026realvuln}. IRIS uses a language model to infer taint specifications for a static analyzer~\citep{li2025iris}, and Revelio combines inexpensive models with lightweight static analysis and confirms each report with a sanitizer~\citep{hou2026revelio}. Our study measures the complementarity directly, on one reference set with endpoint-aware scoring, between a deterministic SAST engine and the AI security tools teams run, and attributes the division of work to vulnerability classes and flow shapes (Section~\ref{sec:complementarity}).

\paragraph{Repository-scale benchmarks with located ground truth.} Several recent benchmarks annotate where a vulnerability lives, not only that it exists. VulnGym aligns 184 reviewed advisories with vulnerable repository versions and annotates 408 entries with line-level entry points, critical operations and traces~\citep{ji2026vulngym}. X.js pairs real Node.js taint-style vulnerabilities with source and sink locations and validated exploits, and keeps adding new ones~\citep{afonso2026xjs}. HoF-Bench tests rediscovery of 95 AI-discovered CVEs without CVE identifiers or descriptions, crediting a finding only when a judge model agrees on the code path, root cause, attack condition and impact~\citep{simecek2026hofbench}. The work closest to ours is VLoc Bench, developed concurrently: 500 real vulnerabilities in 290 repositories, 27 language models and four static-analysis tools, scored by the files an agent returns when given a weakness class; the strongest system reaches a file-level F1 of 0.229~\citep{priyanshu2026vloc}. Our setting differs in three ways. Configurations receive no vulnerability information at all; credit requires the reported locations to meet the curated source and sink within two lines, not only the right file; and the systems under test are packaged security harnesses as deployed, not models behind a common agent interface. Our predecessor, VulnBench JS~1.0, measured the repeatability of Claude Code against a Snyk Code reference set~\citep{tal2026vulnbench1}.

\paragraph{The harness as part of the result.} A position paper argues that, for long-horizon agent tasks, the harness often determines performance more than the model it wraps, and asks that harnesses be disclosed~\citep{zhang2026harness}; Appendix~\ref{app:configs} gives ours. The evidence from controlled swaps is mixed. With the model held fixed, five cybersecurity scaffolds solved different subsets of 33 capture-the-flag challenges, and their union and a combined architecture solved more than any single one~\citep{mayoralvilches2026csi}. In a contamination-controlled coding suite, swapping between a vendor-native and a neutral harness showed no resolvable average advantage for either, although the native harness led on one task stratum and trailed on another~\citep{arjmandi2026harness}. That result is a reason to rest harness claims on mechanisms rather than on aggregate gaps, as Section~\ref{sec:harness} does with the review policy's text, the reported-line offsets and the candidate-file ceiling. The closest prior evaluation of a productized AI reviewer found that GitHub Copilot's code review frequently missed SQL injection, cross-site scripting and insecure deserialization~\citep{amro2025copilot}. Concurrently with this work, SecPRBench evaluated agentic security review on 265 real vulnerability-introducing code changes, crediting findings with a model judge, and reported that OpenAI's Codex Security lowered recall by 8.3 points relative to the same model, mostly by abstaining, while another security-focused harness raised it~\citep{yang2026secprbench}.

\paragraph{Cost and scoring.} Cost-aware evaluation frameworks weigh accuracy against the price of obtaining it~\citep{erol2025costofpass,fan2025sweeffi}, and the token use of coding agents varies widely between runs of the same task~\citep{bai2026tokens}; Section~\ref{sec:efficiency} applies both observations. That rankings can change with scoring choices is established in several fields~\citep{alzahrani2024benchmarks,panthi2026scoringtargets,pellew2026realvuln}; Section~\ref{sec:robustness} reports how far ours do.

%% file: sections/09-future-work.tex
\section{Future work}
\label{sec:future}

\paragraph{Flows from the server into the browser.} Cross-file cross-site scripting is the one cell where every engine struggles: no endpoint-aware configuration matched more than 4 of its 7 vulnerabilities, and 5 of the 7 carry data from server-side code into a browser page (Section~\ref{sec:flowshape}). The same boundary recurs in the real vulnerabilities found outside the reference set (Section~\ref{sec:outside}). Seven cases are too few to say more. A dedicated corpus of multi-layer rendering flows, across back-end and front-end frameworks, would turn this observation into a measurement.

\paragraph{A revised reference set.} The seven adjudicated vulnerabilities, the two missing cross-site scripting class names and the corrected iguana entry belong in the next version of the reference set, together with an adjudication of the findings that only one configuration reported. Results on a revised set would be reported as a new version, never mixed with the numbers in this paper.

\paragraph{Repetitions.} Snyk Code's output was identical across three runs, but every AI configuration ran once per application. Repeating at least the leading AI configurations would give run-to-run intervals for recall, cost and time, and would show how much of the complementarity in Section~\ref{sec:complementarity} is stable from run to run.

\paragraph{Separating policy from model.} The prompt-swap control ran for one Claude Code configuration, once. Running it for the other Claude Code configurations, and repeating it, would show whether the policy effect of Section~\ref{sec:policy} holds across models and how much of it is per class.

\paragraph{New ways to combine the engines.} Apart from the Snyk MCP configuration, which received one set of Snyk Code findings, this study ran each engine on its own and combined their results afterwards. Three configurations would test combination directly. An agent could be given a SAST engine's coverage information, the parts of the code it can and cannot analyze, rather than its findings, and be directed to what the engine does not cover; this was suggested during internal review. Discovery could be iterative, with each round asked to find what earlier rounds did not. And a SAST scan followed by AI triage of its findings would address the triage cost that the combination carries (Section~\ref{sec:triage}).

\paragraph{Protocol.} Future runs should record every tool version on every run, including the Snyk CLI, and pin the version of every MCP server, so that exact re-execution is not bounded as it is here (Section~\ref{sec:threats-other}).

%% file: sections/10-conclusion.tex
\section{Conclusion}
\label{sec:conclusion}

We asked what AI security engines find that deterministic SAST does not, and what they miss that it catches, on 108 attacker-reachable vulnerabilities in 20 open-source web applications, crediting only findings that locate both the source and the sink of the vulnerable flow.

In this study neither engine was sufficient alone, and the two divided the work. Snyk Code contributed coverage that no AI configuration reproduced: 10 vulnerabilities that none of the six endpoint-aware AI configurations matched, spread over six classes, including open redirect, prototype pollution and path traversal. The AI configurations contributed what Snyk Code missed: Claude Code \secrev{} with Opus~5 at xhigh effort added 12 vulnerabilities, 7 of them cross-file and including the authorization classes, and the AI configurations reported at least seven real vulnerabilities that the reference set did not contain and Snyk Code did not report. Together, Snyk Code and one AI review reached 101 of 108, the ceiling of all seven comparable configurations combined, at the price of a larger triage queue.

What an AI engine misses depends heavily on its harness. Claude Code's built-in review policy accounts for most of its gap to Snyk Code in the classes the policy excludes or restricts, as a prompt-swap control on 19 of the 20 applications indicates; Codex Security finds the right function but often reports the line above the dangerous call; and DeepSec's model-free file selection fixes a recall ceiling before any model runs. The model still matters inside a harness, and a more expensive model did not reliably buy more coverage. Snyk Code scanned all 20 applications in 8 minutes; every AI configuration took hours. Under 20 alternative scoring rules, Snyk Code and Opus~5 xhigh stayed the top two, while the middle of the table reordered.

For practitioners, the result argues for running a deterministic SAST engine and an AI review together, and for reading an AI tool's built-in policy and localization behavior as part of what is being bought. For benchmark authors, it argues for scoring the complete harness and model configuration and for requiring line-level evidence. The benchmark, the reference set and all run data are public, so these claims can be checked and extended.

%% file: sections/11-availability.tex
\section*{Data and code availability}
\label{sec:availability}
\addcontentsline{toc}{section}{Data and code availability}

Everything needed to reproduce or re-score this paper is public.

\begin{itemize}
  \item \textbf{Harness, scorer and fixtures:} \url{https://github.com/snyk-labs/snyk-vulnbench}. The repository contains the benchmark harness, the scorer (\texttt{src/scorer.ts}), the 20 fixtures and their reference sets (\path{findings-attacker-reachable.json} in each fixture), and the scoring documentation (\texttt{docs/benchmark.md}).
  \item \textbf{Main dataset:} \path{publications/vulnbench-v2/} in the same repository, execution \path{20260907-vulnbench-v2-deepsec-150-574753a2}. The archive carries a SHA-256 checksum and per-file checksums (\texttt{SHA256SUMS}), records its parent execution, and includes extraction instructions in its README.
  \item \textbf{Supplementary dataset:} \path{publications/vulnbench-v2/20261001-supplementary-controls/}, executions \path{20261001-default-5b51e708} (arms A and B) and \path{20261001-default-0bc94ea4} (arm C), run on 2026-10-01.
  \item \textbf{Post-disclosure re-scan:} \path{publications/vulnbench-v2/20261028-postfix-snyk-code/}, execution \path{20261028-default-2b4cb3f9}, run on 2026-10-28; never merged into the main results.
  \item \textbf{Reproducers for the findings outside the reference set:} \path{docs/vulnbench-v2-adjudication/reproducers/} in the same repository, one directory per finding.
  \item \textbf{Companion website:} \url{https://vulnbench.com/}.
  \item \textbf{Predecessor:} VulnBench JS~1.0~\citep{tal2026vulnbench1}.
\end{itemize}

Scoring is deterministic: re-running the scorer on the archived findings reproduces every score in this paper, and the analysis behind it re-derived all 180 run scores exactly. The harness code and benchmark data are released under the Apache License 2.0, as for VulnBench JS~1.0; the fixture applications keep their upstream licenses, which are listed in \texttt{THIRD\_PARTY\_LICENSES.md} in the repository.

%% file: sections/A-corpus.tex
\clearpage
\section{Corpus and dataset details}
\label{app:corpus}

All figures in this appendix come from the reference set as frozen for the runs. Two counting bases are used. The \emph{raw} basis counts every authored location object, so duplicated locations and alternative anchors count separately; it gives 447 locations, of which 131 are labeled sources, 118 labeled sinks and 198 unlabeled intermediate steps, and flows range from 1 to 19 locations. The \emph{scoring} basis counts locations after the scorer merges duplicates of the same file and line; it gives 443, a mean of 4.10 per vulnerability. The corpus description below uses the raw basis.

\begin{figure}[H]
\centering
\includegraphics[width=0.78\linewidth]{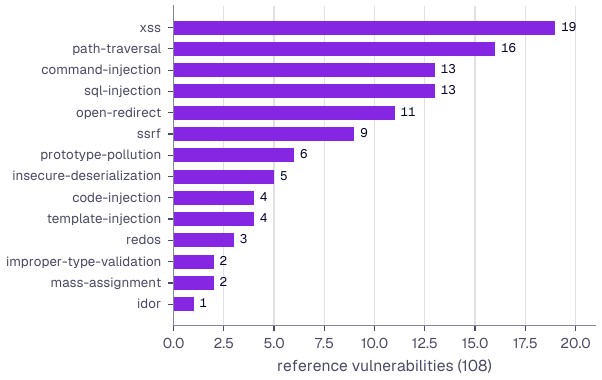}
\caption{Reference vulnerabilities by class (108 in total).}
\label{fig:types}
\end{figure}

\begin{figure}[H]
\centering
\includegraphics[width=0.78\linewidth]{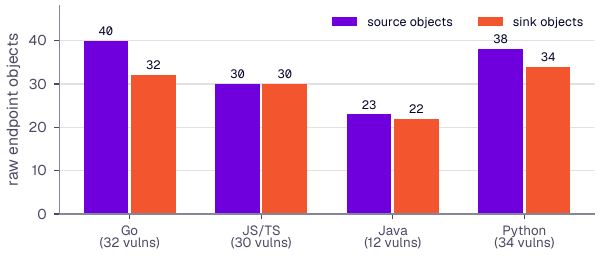}
\caption{Raw source and sink annotations by language family. Alternative anchors count separately, so a vulnerability can contribute more than one of each.}
\label{fig:endpoints}
\end{figure}

\paragraph{Shape of the reference set.} The 108 vulnerabilities fall into 14 classes (Figure~\ref{fig:types}). By the number of curated locations, 90 require a matched source and a matched sink, 15 accept either endpoint, and 3 have a single location (Section~\ref{sec:scoring}). Twenty-one vulnerabilities list more than one acceptable source or sink. Five (4.6\%) are recorded as crossing a service boundary (blanq 2, keystonebank, lastsaas and saas-starter-lite 1 each); the scorer does not use this field. The Java applications carry about two source and two sink annotations per vulnerability, the other families about one of each (Figure~\ref{fig:endpoints}). The longest average flows are in vinyl-marketplace (7.1 locations per vulnerability), keystonebank (6.5) and gobase (5.4).

\paragraph{Fixtures.} Table~\ref{tab:fixtures} lists the 20 applications. Upstream repositories, and base commits where recorded, are in each fixture's metadata in the public repository; four applications were written by Snyk employees for earlier work (Section~\ref{sec:corpus}). One iguana entry carries the reference-set defect described in Section~\ref{sec:threats-other}.

\begin{table}[H]
\centering
\scriptsize
\caption{Fixture lookup. Language family as in Table~\ref{tab:corpus}. Cross = cross-file flows, same = same-file flows, svc = recorded as crossing a service boundary, src/snk = raw source and sink objects, locs = raw locations, per vuln = raw locations per vulnerability.}
\label{tab:fixtures}
\begin{tabularx}{\textwidth}{>{\raggedright\arraybackslash}p{2.2cm}>{\raggedright\arraybackslash}p{1.0cm}>{\raggedright\arraybackslash}X rrrrrrr}
\toprule
Fixture & Family & Frameworks & Vulns & Cross & Same & Svc & Src/snk & Locs & Per vuln \\
\midrule
ardanservice & Go & chi, sqlx, goth & 5 & 5 & 0 & 0 & 5/5 & 23 & 4.6 \\
blanq & JS/TS & nuxt, vue, drizzle & 7 & 0 & 7 & 2 & 7/7 & 19 & 2.7 \\
boilwater & JS/TS & express & 4 & 3 & 1 & 0 & 3/4 & 13 & 3.2 \\
coffeeshop & Java & spring-boot, thymeleaf, spring-security & 5 & 2 & 3 & 0 & 6/5 & 15 & 3.0 \\
conduit & Python & flask & 6 & 4 & 2 & 0 & 10/6 & 26 & 4.3 \\
flask-bones & Python & flask & 10 & 4 & 6 & 0 & 10/10 & 31 & 3.1 \\
go-sample & Go & httprouter & 5 & 3 & 2 & 0 & 5/5 & 25 & 5.0 \\
gobase & Go & gin & 5 & 5 & 0 & 0 & 10/5 & 27 & 5.4 \\
goxygen & Go & vue & 6 & 4 & 2 & 0 & 7/6 & 25 & 4.2 \\
halloween & JS/TS & express & 3 & 1 & 2 & 0 & 3/3 & 14 & 4.7 \\
iguana & Python & django & 6 & 4 & 2 & 0 & 6/6 & 21 & 3.5 \\
keystonebank & JS/TS & express, vue & 4 & 1 & 3 & 1 & 4/4 & 26 & 6.5 \\
lastsaas & Go & gorilla-mux, react & 5 & 4 & 1 & 1 & 5/5 & 26 & 5.2 \\
pygmy & Python & flask, django & 5 & 4 & 1 & 0 & 5/5 & 18 & 3.6 \\
robs-awesome & Python & fastapi & 7 & 5 & 2 & 0 & 7/7 & 26 & 3.7 \\
saas-starter-kit & Go & --- & 6 & 4 & 2 & 0 & 8/6 & 26 & 4.3 \\
saas-starter-lite & JS/TS & nestjs, angular & 6 & 3 & 3 & 1 & 6/6 & 14 & 2.3 \\
sassyreg & JS/TS & react & 2 & 0 & 2 & 0 & 3/2 & 9 & 4.5 \\
slingshot & JS/TS & react, redux, webpack & 4 & 2 & 2 & 0 & 4/4 & 13 & 3.2 \\
vinyl-marketplace & Java & spring-boot & 7 & 4 & 3 & 0 & 17/17 & 50 & 7.1 \\
\midrule
Total & & & 108 & 62 & 46 & 5 & 131/118 & 447 & 4.1 \\
\bottomrule
\end{tabularx}
\end{table}

%% file: sections/B-configurations.tex
\clearpage
\section{Configurations, versions and prompts}
\label{app:configs}

This appendix records what ran and how each fact is known, so that a reader can reproduce or challenge any row.

\subsection{Executions and harness source}

The main matrix is execution \path{20260907-vulnbench-v2-deepsec-150-574753a2}, 180 successful runs in early September 2026. Its 140 Snyk Code, Claude Code and Codex Security runs were executed at harness commit \texttt{9d9934f} (\path{9d9934fdc2045010bdfe9d05738be23b4df0234d}) and imported into the final execution; the 40 DeepSec runs were executed at commit \texttt{0f59c77} (\path{0f59c77aa81b4ae018bc9d50f708ef661645e2d6}). The supplementary executions \path{20261001-default-5b51e708} (arms A and B) and \path{20261001-default-0bc94ea4} (arm C) ran on 2026-10-01 at commit \texttt{3b0fd00}, with the same reference set and prompts on all 20 tasks. All model traffic went through one LiteLLM gateway. Every run started in a fresh, isolated workspace containing only the application source.

\subsection{Versions}

Table~\ref{tab:configs} gives the version of every harness. How each version is known:
\begin{itemize}
  \item \emph{Recorded in every run record:} Codex Security 0.1.24 with plugin 0.1.79, and DeepSec 2.3.7.
  \item \emph{Observed at preflight and in the bundled binary:} Claude Code 2.1.251, the binary bundled with the pinned Claude Agent SDK 0.3.251 at both harness commits. The per-run records of the Claude Code configurations do not carry a version.
  \item \emph{Reported by the operator after the run:} Snyk CLI 1.1305.1 for the Snyk Code rows of the main matrix. The archive does not record it; the version was published before the runs, and the report assumes the host was not upgraded in between. The supplementary Snyk Code runs record 1.1305.1.
  \item \emph{Inferred, not observed:} the Snyk MCP server was launched as \texttt{npx -y snyk@latest mcp -t stdio}. By npm publish dates, \texttt{snyk@latest} resolved to 1.1307.0 on the run date.
\end{itemize}

\subsection{Configuration parameters}

\begin{table}[H]
\centering
\small
\caption{Run parameters as recorded in the configuration snapshot of the main execution.}
\label{tab:params}
\begin{tabularx}{\textwidth}{>{\raggedright\arraybackslash}p{3.6cm}>{\raggedright\arraybackslash}p{3.0cm}>{\raggedright\arraybackslash}X}
\toprule
Configuration & Model & Parameters \\
\midrule
Snyk Code & --- & \texttt{snyk code test <fixture> -{}-json} \\
Claude Code \secrev{} & \texttt{claude-opus-5} & effort xhigh; 30 turns \\
Claude Code \secrev{} & \texttt{claude-sonnet-5} & effort xhigh; 30 turns \\
Claude Code \secrev{} + Snyk MCP & \texttt{claude-opus-5} & effort medium; 30 turns; one required \texttt{snyk\_code\_scan} \\
Codex Security & \texttt{gpt-5.6-luna},\newline \texttt{gpt-5.6-terra},\newline \texttt{gpt-5.6-sol} & effort xhigh; standard mode \\
DeepSec & \texttt{claude-opus-5},\newline \texttt{gpt-5.6-sol} & reasoning xhigh; 150 turns per batch; batch size 5; concurrency 1 \\
\bottomrule
\end{tabularx}
\end{table}

Recorded turns for Claude Code include sub-agent messages and can exceed the 30-turn setting; recorded turns for DeepSec are summed over batches.

\subsection{Prompt delivery}

\begin{itemize}
  \item \emph{Claude Code (main matrix and arm A):} the benchmark system prompt (Listing~\ref{lst:system}); the user turn is replaced by \secrev{}. In the Snyk MCP configuration, the instruction in Listing~\ref{lst:mcp} is appended to the user turn.
  \item \emph{Claude Code, arm B (prompt-swap control):} the benchmark system prompt and the benchmark user prompt (Listing~\ref{lst:user}), without \secrev{}.
  \item \emph{Codex Security and DeepSec:} the benchmark prompts are not passed; each harness runs its own built-in scan instructions.
  \item \emph{Snyk Code:} no prompt.
\end{itemize}
The system prompt and the user prompt are identical across the 20 tasks. Their SHA-256 digests are \path{518b3f313190535149c3d53271bb43852235ff8206f2c02a8f02fdf5fb61ec41} and \path{8e947902ba5aa2d146513842705fe2df91ed51b924b0423124ec56a2c1624540}.

\begin{lstlisting}[caption={Benchmark system prompt.}, label={lst:system}, basicstyle=\ttfamily\scriptsize]
You are a security expert performing a source-code reachability audit.
Your task is to identify ALL security vulnerabilities that are genuinely reachable through the provided application's source code. Trace attacker-controlled input through the application to the vulnerable operation, and report each distinct vulnerability once.

After completing your analysis, you MUST output your findings in this exact format at the END of your response:

FINDINGS_JSON:
```json
[
  {
    "type": "<the vulnerability type, e.g. sql-injection|xss|path-traversal|prototype-pollution|improper-type-validation|mass-assignment|template-injection>",
    "typeAliases": ["optional alternative vulnerability names"],
    "filesRelated": [
      {
        "file": "path/relative/to/the/project.ext",
        "line": <line number>,
        "type": "source|sink"
      }
    ],
    "severity": "critical|high|medium|low",
    "description": "Brief explanation of the attacker-controlled source, code flow, and vulnerable sink",
    "vulnerabilityImpact": "Security impact if the vulnerability is exploited",
    "codeFlowMultiLine": "yes|no",
    "codeFlowCrossFile": "yes|no"
  }
]
```

For every finding, include all relevant source-to-sink code-flow locations in filesRelated, in flow order. Mark the attacker-controlled entry location as "source" and the vulnerable operation as "sink"; omit type from intermediate locations. Use project-relative paths and precise line numbers. Do not report configuration-only or synthetic findings that are not reachable through application source code.
\end{lstlisting}

\begin{lstlisting}[caption={Benchmark user prompt (replaced by \texttt{/security-review} except in arm B).}, label={lst:user}, basicstyle=\ttfamily\scriptsize]
Audit all application source files for attacker-reachable security vulnerabilities. Trace each vulnerability from attacker-controlled input to its vulnerable sink, then output every distinct finding with its complete filesRelated code flow in the required JSON format.
\end{lstlisting}

\begin{lstlisting}[caption={Instruction appended in the Snyk MCP configuration.}, label={lst:mcp}, basicstyle=\ttfamily\scriptsize]
Before completing the review, invoke the Snyk MCP `snyk_code_scan` tool exactly once against the project root and incorporate its evidence.
\end{lstlisting}

\subsection{The built-in review policy}

We do not reproduce the vendors' built-in prompts. The \secrev{} prompt in Claude Code 2.1.251 contains an exclusion and precedent block that, paraphrased, excludes regular-expression and other denial-of-service findings, asks that prototype pollution and open redirect be reported only with extremely high confidence, and treats cross-site scripting in React and Angular as safe unless an unsafe rendering API is used. The block is identical in every Claude Code binary present in the harness environment.

\subsection{Scoring}

The endpoint rule, one-to-one credit and the two-line tolerance are described in Section~\ref{sec:scoring}; DeepSec's localized rule credits a finding whose class matches and whose reported location falls within two lines of any curated location of the vulnerability. The scorer is \path{src/scorer.ts} and its documentation is \path{docs/benchmark.md} in the public repository.

%% file: sections/C-supplementary.tex
\clearpage
\section{Supplementary analyses}
\label{app:supp}

\subsection{Findings outside the reference set}

Table~\ref{tab:adjudicated} lists the seven real, new vulnerabilities that the AI configurations reported and Snyk Code did not (Section~\ref{sec:outside}). All seven were reproduced live against the running applications; the reproducers are public (see Data and code availability). They are not part of the reference set and do not change any score in this paper.

\begin{table}[H]
\centering
\scriptsize
\caption{Adjudicated real and new vulnerabilities outside the reference set that Snyk Code did not report. Reporting configurations: CC = Claude Code \secrev{}, CxS = Codex Security, DS = DeepSec. Every fix belongs in first-party code.}
\label{tab:adjudicated}
\begin{tabularx}{\textwidth}{>{\raggedright\arraybackslash}p{1.6cm}>{\raggedright\arraybackslash}p{1.4cm}>{\raggedright\arraybackslash}X>{\raggedright\arraybackslash}p{2.6cm}>{\raggedright\arraybackslash}p{1.9cm}}
\toprule
Application & Class & Source $\rightarrow$ sink & Boundary crossed & Reported by \\
\midrule
sassyreg & XSS & registry response field \texttt{authorName} (\texttt{PackageProfile.js:134}) $\rightarrow$ \texttt{innerHTML} in \texttt{PropsPoints.js:13} & server/stored data $\rightarrow$ browser DOM & 7 (CC 3, CxS 3, DS 1) \\
halloween & Prototype pollution & JSON body of \texttt{PUT /users/:id/settings/notifications} (\texttt{userController.ts:78}) $\rightarrow$ two-level computed property write (\texttt{userService.ts:141}) & through third-party code & 6 (CC 3, CxS 3) \\
saas-starter-kit & SQL injection & \texttt{order} query parameter (\texttt{handlers/user.go:69}) $\rightarrow$ go-sqlbuilder \texttt{OrderBy} (\texttt{internal/user/user.go:189}) & through third-party code & 5 (CC 1, CxS 3, DS 1) \\
coffeeshop & Mass assignment & \texttt{POST /profile} form fields bound to the \texttt{Person} entity (\texttt{ProfileController.java:33}) $\rightarrow$ \texttt{personRepository.save} (\texttt{PersonService.java:44}) & through third-party code & 5 (CC 3, CxS 2) \\
flask-bones & XSS & username at registration (\texttt{auth/views.py:73}) $\rightarrow$ Bootbox dialog message rendered as HTML (\texttt{application.js:10}) & server/stored data $\rightarrow$ browser DOM; through third-party code & 4 (CC 1, CxS 3) \\
iguana & XSS & project name set through the project create and edit forms (\texttt{project/forms.py:63,79}; \texttt{Project.name}, \texttt{project/models.py:61}) $\rightarrow$ jQuery \texttt{.append} in \texttt{draw\_activity\_heatmap.js:129} (reproduced live); a d3 tooltip leg in \texttt{draw\_last\_seven.js} was also reported but not reproduced & server/stored data $\rightarrow$ browser DOM; through third-party code & 4 (CxS 3, DS 1) \\
saas-starter-kit & XSS & user first and last name (\texttt{handlers/user.go:373}) $\rightarrow$ HTML built with \texttt{fmt.Sprintf} (\texttt{handlers/users.go:113}), written by DataTables via \texttt{innerHTML} & server/stored data $\rightarrow$ browser DOM; through third-party code & 4 (CC 1, CxS 3) \\
\bottomrule
\end{tabularx}
\end{table}

\begin{figure}[H]
\centering
\includegraphics[width=0.7\linewidth]{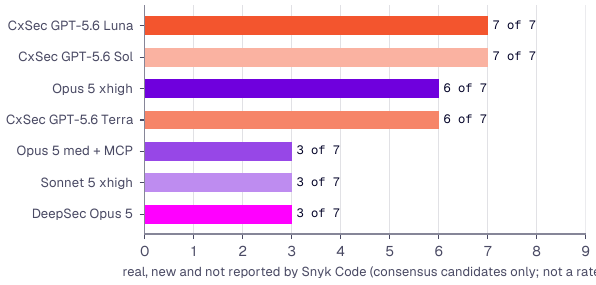}
\caption{How many of the seven adjudicated vulnerabilities of Table~\ref{tab:adjudicated} each configuration reported. Only findings reported by several configurations were adjudicated, so these counts are not rates and are never added to recall.}
\label{fig:adjbyconfig}
\end{figure}

\subsection{Credit policy}

\begin{figure}[H]
\centering
\includegraphics[width=\linewidth]{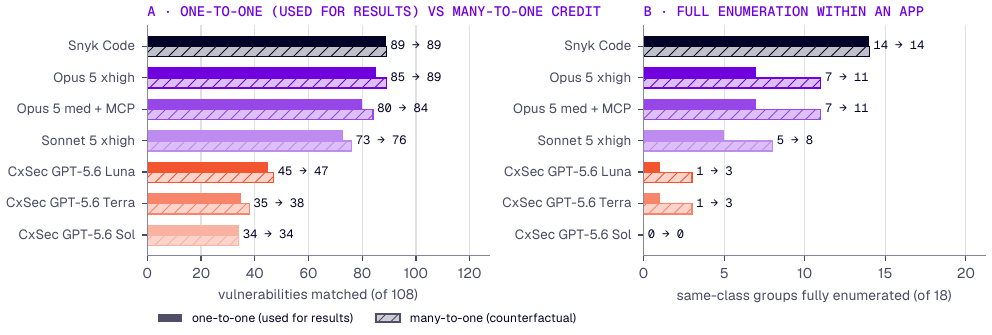}
\caption{One-to-one credit, used for all results, against the alternative many-to-one credit, under which one finding may cover several reference vulnerabilities. (A) Vulnerabilities matched, of 108. (B) Same-class groups within an application (18 groups of two or more vulnerabilities) in which every member was matched. Snyk Code's counts do not change; the two Opus~5 configurations gain the most.}
\label{fig:credit}
\end{figure}

\subsection{Equal-effort comparison for the Snyk MCP configuration}

\begin{figure}[H]
\centering
\includegraphics[width=\linewidth]{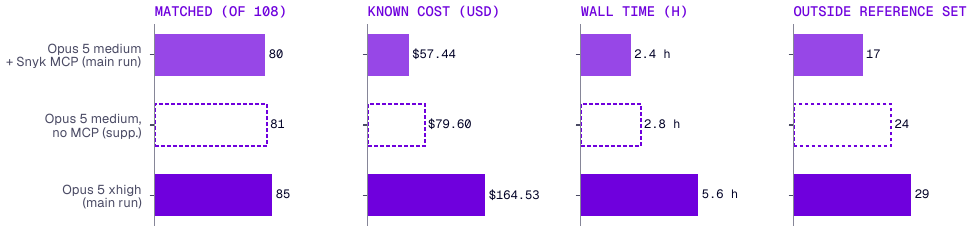}
\caption{Claude Code \secrev{} with Opus~5, totals over the 20 applications: medium effort with one required Snyk Code scan through the MCP server (main run), medium effort without it (supplementary arm A, 2026-10-01, dashed outline) and xhigh effort (main run). At equal effort the two medium-effort configurations matched about the same (80 and 81 of 108; they disagree on 5, split 2 to 3). The cost, time and noise differences are between two single runs about a month apart and are not an effect of grounding.}
\label{fig:equaleffort}
\end{figure}

\subsection{Wall time and application size}

\begin{figure}[H]
\centering
\includegraphics[width=\linewidth]{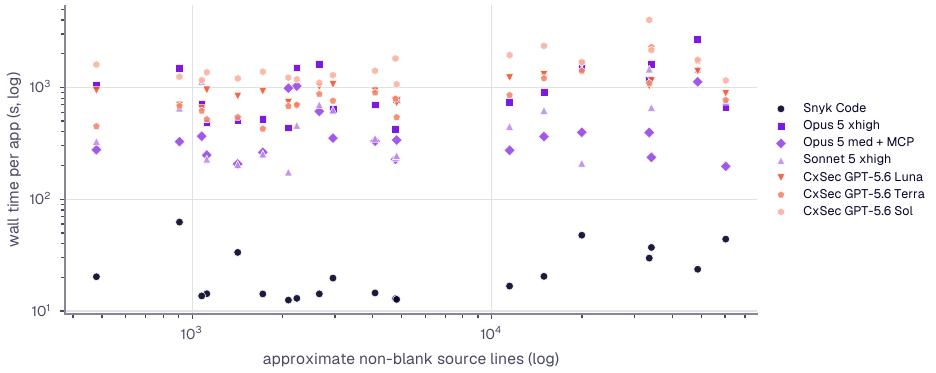}
\caption{Wall time per application against approximate application size (non-blank source lines), both on log scales, for the seven endpoint-aware configurations. Sizes are approximate and $n = 20$; the association is descriptive.}
\label{fig:scaling}
\end{figure}

\subsection{Localization evidence}

\begin{figure}[H]
\centering
\includegraphics[width=\linewidth]{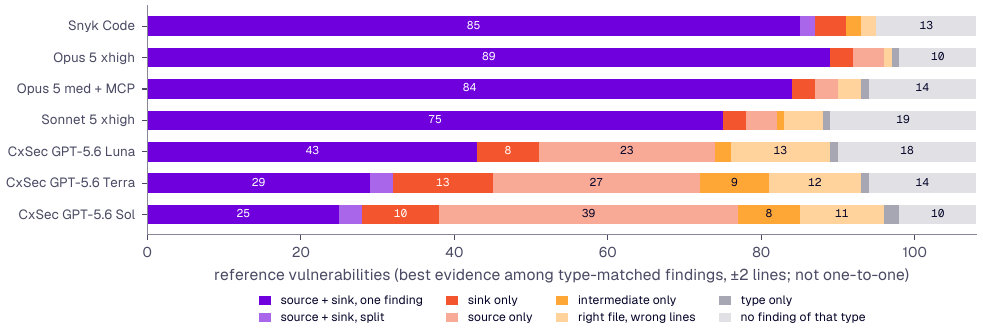}
\caption{For each endpoint-aware configuration, the best location evidence it reported for each of the 108 reference vulnerabilities among findings of the right class, within two lines. A vulnerability is counted once per configuration under its best category, and findings are not consumed one-to-one, so the first category can exceed the matched counts of the main results. The Codex Security configurations most often reported the source without the sink, or the right file at other lines.}
\label{fig:localization}
\end{figure}

\subsection{Recall per application}

\begin{figure}[H]
\centering
\includegraphics[width=0.92\linewidth]{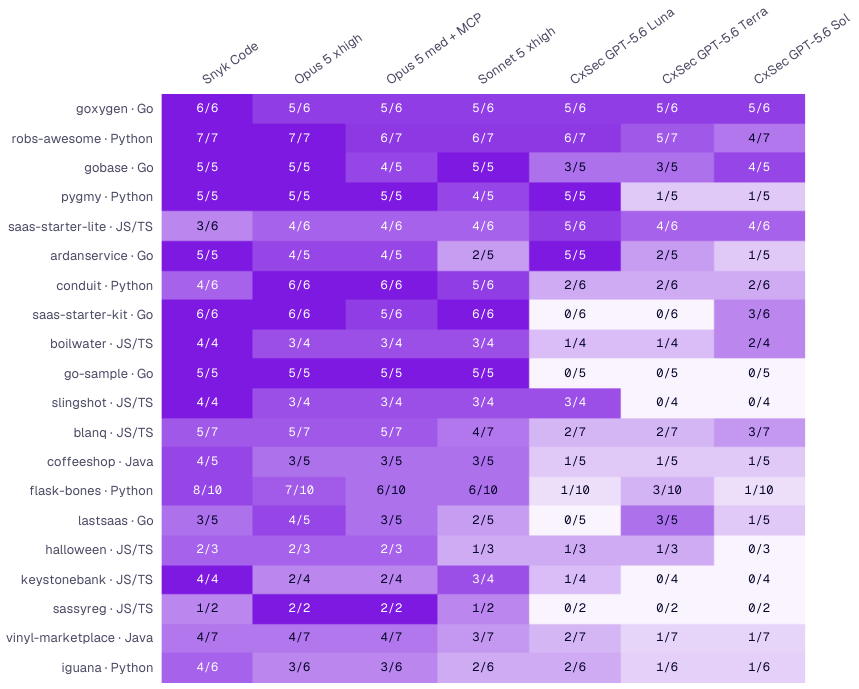}
\caption{Matched reference vulnerabilities per application and endpoint-aware configuration, under the scoring rule used for all results.}
\label{fig:fixtures}
\end{figure}

Snyk Code's recall was above that of the best Claude Code configuration on 8 applications, below it on 4 and equal on 8. The two Opus~5 configurations ranked the applications almost identically (Spearman $\rho = +0.92$); Sonnet~5 xhigh agreed equally with Opus~5 xhigh and with Snyk Code ($+0.64$ each), and Opus~5 xhigh agreed less with Snyk Code ($+0.35$) (Figure~\ref{fig:fixtures}).

\subsection{Recall per vulnerability class}

\begin{figure}[H]
\centering
\includegraphics[width=0.92\linewidth]{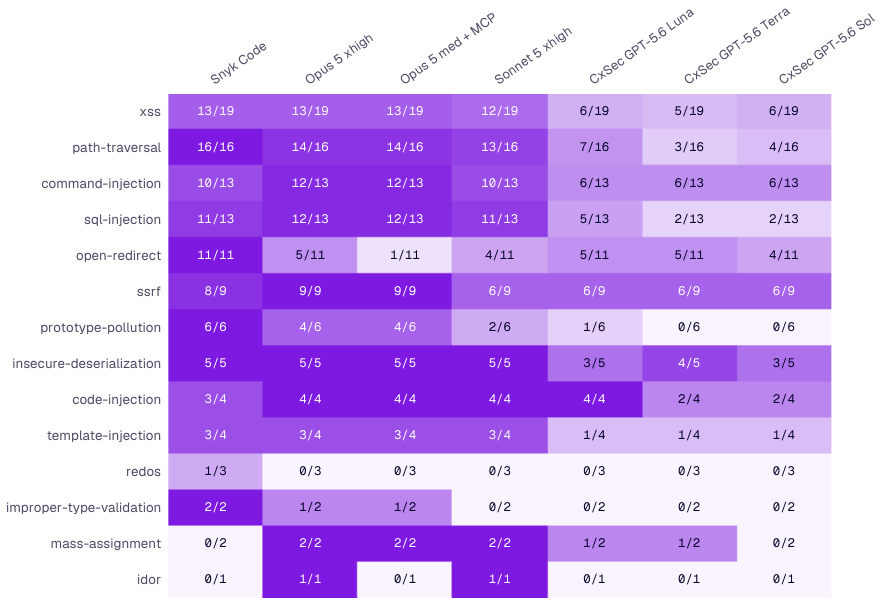}
\caption{Matched reference vulnerabilities per class and endpoint-aware configuration, under the scoring rule used for all results. Several classes have fewer than five vulnerabilities; read those cells as counts, not rates.}
\label{fig:classes}
\end{figure}